\documentclass[11pt]{article}

\usepackage[T1]{fontenc}
\usepackage[utf8]{inputenc}
\usepackage{lmodern}
\usepackage{geometry}
\usepackage{setspace}
\usepackage{graphicx}
\usepackage{booktabs}
\usepackage{array}
\usepackage{longtable}

\usepackage{caption}
\usepackage{subcaption}

\usepackage{amsmath}
\usepackage{amssymb}

\usepackage{hyperref}
\hypersetup{
    colorlinks=true,
    linkcolor=black,
    citecolor=black,
    urlcolor=black
}

\title{\textbf{Reimagining the Traditional Flight Computer:\\
E6BJA as a Modern, Multi-Platform Tool for Flight Calculations and Training}}

\author{
Dr Jamie J. Alnasir \\
\small Researcher, Software Engineer and UK Licensed Private Pilot\\
\small \texttt{jamies.aviation@al-nasir.com}
}

\date{4th January 2026 (v2)}

\begin{document}

\maketitle

%  --------------------
% Abstract
%  --------------------
\begin{abstract}
Traditional flight computers --- including mechanical “whiz-wheels” (e.g.\ E6B, CRP series) and electronic flight calculators (e.g.\ ASA CX-3, Sporty’s E6-B) --- have long played a central role in flight planning and training within general aviation (GA). While these tools remain pedagogically valuable, their fixed form factors, constrained interaction models, and limited extensibility are increasingly misaligned with the expectations and workflows of pilots operating in modern digital environments.\newline
This paper presents E6BJA (Jamie’s Flight Computer), a fully featured, multi-platform, software-based flight computer designed natively for Apple iOS, Android, and Microsoft Windows devices, with a complementary web-based implementation. E6BJA reproduces the core calculations of traditional flight computers while extending them through enhanced modelling capabilities and more accurate atmospheric (i.e.  ISA-based) and performance calculations — including carburettor icing risk estimation and aircraft-specific weight and balance modelling for common GA aircraft.  Each calculator is accompanied by embedded educational monographs that explain underlying assumptions, variables, and equations.\newline
We compare E6BJA with mechanical and electronic flight computers across functional, cognitive, and technical dimensions, demonstrating improvements in accuracy, error reduction, discoverability, and educational value. We also discuss key design trade-offs associated with native multi-platform development and examine how contemporary mobile computing environments can support safer and more intuitive pre-flight planning for pilots, trainees, instructors, and flight planning personnel.\newline
By combining the conceptual rigour of traditional flight planning methods with modern human--computer interaction design, E6BJA represents a meaningful evolution in pilot-facing flight tools. Its integration of computation with contextual explanation enables a dual role as both calculation aid and instructional resource; for example, illustrating the relationship between bank angle and stall speed in the Load Factor calculator using cosine-based logic. This approach provides particular value in training and instructional contexts, supporting deeper conceptual reinforcement during pre- and post-flight briefing.
\end{abstract}


%  --------------------
\section{Introduction}

Flight computers have played a foundational role in general aviation (GA) for over half a century. Whether used in the classroom, in the cockpit, or during flight planning sessions, these tools serve to reinforce fundamental aeronautical knowledge and, in certain scenarios, can support real-time decision-making. Traditionally, this role has been fulfilled by mechanical slide-rule devices, most notably the E6B and CRP series “whiz-wheels”, which have been widely adopted by flight training organisations worldwide. These devices have served as standard tools in aviation education and dead reckoning since the early twentieth century. The E 6B itself, originally developed by Philip Dalton and adopted by the U.S. Army Air Corps in the late 1930s, established a durable standard for pilot navigation training, and its fundamental design principles continue to underpin flight computer instruction today \cite{Sanik1997} \cite{vanRiet2007}.

\subsection*{Contributions}
This paper makes the following contributions:
\begin{itemize}
  \item A structured analysis of mechanical and electronic flight computers, including characteristic error profiles and cognitive limitations arising from manual interpolation, fixed-function hardware, and constrained interaction models.
  \item The design, implementation, and architectural evaluation of \emph{E6BJA (aka Jamie's Flight Computer)}, a native, multi-platform, offline-capable software-based flight computer that integrates computation, visualisation, and embedded educational support.
  \item A comparative evaluation across functional, cognitive, and technical dimensions, demonstrating how modern software-based flight computers can improve usability, error transparency, and pedagogical value while preserving the conceptual rigour of traditional flight planning methods.
\end{itemize}

\subsection{Mechanical Flight Computers: Principles and Error Profiles}
Mechanical flight computers such as the E6B and CRP series whiz-wheels are based on analogue circular slide rule principles, combining logarithmic scales with geometric constructions to solve common aeronautical problems. The CRP 1, for example, consists of a fixed outer scale, a rotating inner scale, and a sliding wind grid, enabling multiplication, division, unit conversion, and vector-based dead reckoning through physical alignment rather than symbolic computation \cite{PooleysCRP1Booklet2017}.\newline

For scalar calculations  ---  such as time, distance, speed, fuel consumption, and unit conversion  ---  the device relies on logarithmic interpolation between printed graduations. Results are therefore approximate by design and require the user to estimate intermediate values visually. Accuracy is dependent on scale resolution, lighting conditions, and the operator’s ability to interpolate between markings, introducing a non-trivial human error component .
Vector-based calculations, including wind triangle solutions, headings, ground speed, and crosswind components, are performed using the sliding wind grid and centre dot. These constructions require the pilot to manually position wind vectors, align track or heading references, and read off results from intersecting scales. Errors can arise from misplacement of the wind vector, incorrect selection of speed arcs, or misreading of angular offsets --- particularly under time pressure or turbulence. The handbook explicitly distinguishes between low speed and high speed ranges, requiring users to select the appropriate scale before solving a problem, further increasing cognitive and procedural load.\newline

Several limitations of the mechanical format give rise to characteristic error profiles. First, values that fall outside the printed scale --- such as very high true airspeeds or extreme wind components  ---  cannot be represented directly and must be approximated or rescaled mentally. Second, chained calculations (e.g. fuel planning combined with climb performance and wind correction) require repeated manual realignment, compounding small interpolation errors at each step. Third, the device provides no explicit indication of intermediate assumptions or constraints, making it difficult for learners to diagnose incorrect results once an error has been introduced.\newline

Despite these limitations, the CRP 1 handbook emphasises that such devices are intentionally non-automated and non-persistent: they store no data, enforce no hidden logic, and require continuous user engagement. This transparency has historically made whiz-wheels effective instructional tools, but it also explains why their accuracy, reproducibility, and scalability are inherently bounded by human perception and manual dexterity.\newline

Despite their longevity, mechanical flight computers are increasingly at odds with the expectations and operational contexts of today’s pilots. Their design demands manual precision and interpretive skill, with limited support for multi-step calculations, unit conversions, or verification of results. While they are pedagogically valuable for reinforcing core mathematical and aerodynamic principles, they are inherently error-prone, cognitively demanding, and unsuitable for dynamic, high-pressure environments.

\subsection{Electronic Flight Calculators}

To address the limitations of mechanical whiz-wheels, electronic flight calculators such as Sporty’s Electronic E6-B and the ASA CX-3 were introduced. Both the Sporty’s E6-B and ASA’s CX-3 trace their lineage back to the mechanical Dalton “whiz-wheel” E-6B, incorporating many of its core computational principles in digital form \cite{Sanik1997}. These battery-powered, handheld devices replicate the core computational functions of traditional mechanical flight computers, including wind correction angle, ground speed, fuel burn, and density-altitude conversions, while offering faster calculation and improved numerical consistency through digital input and display.\newline

Both devices are explicitly authorised for use during FAA and Canadian aviation knowledge examinations, as stated in manufacturer documentation and consistent with published examination and EFB (Electronic Flight Bag) policy guidance \cite{SportysE6BManual2016,Babb2017}. This authorisation is limited to written examination settings and does not constitute operational approval, certification, or regulatory endorsement for in-flight use.\newline

Architecturally, electronic flight calculators represent an incremental rather than transformative evolution. They are implemented as dedicated, menu-driven hardware systems with constrained displays and tightly bounded software architectures. While the ASA CX-3 supports firmware updates and non-volatile storage of selected planning data, its functionality remains limited to a manufacturer-defined feature set and interaction model. As a result, neither device readily supports the introduction of new calculator paradigms, advanced visualisation tools, or pedagogically rich explanatory layers comparable to those achievable in modern software-based flight computer applications.\newline

These devices are also inherently device-bound and operate in isolation from broader digital ecosystems. They lack integration with mobile platforms, cloud-based planning tools, or cross-device workflows that increasingly characterise contemporary pilot operations. Their deliberately self-contained design  ---  excluding internet connectivity, programmable extensibility, and aircraft-specific configuration  ---  has nonetheless contributed to their widespread adoption in training environments \cite{SportysE6BManual2016}, where predictability, standardisation, and exam compliance are prioritised.\newline

Consequently, while electronic flight calculators remain operationally sufficient for standardised examination use, they fall short in supporting pedagogical scaffolding, extensibility beyond firmware-level updates, or aircraft-specific performance modelling. For example, neither the ASA CX-3 nor Sporty’s Electronic E6-B allows configuration of aircraft-specific weight and balance data, requiring pilots and students to rely on separate charts or paper-based POH (Pilot Operating Handbook) references for such planning tasks.

\subsection{E6BJA Software-Based Flight Computer Application}

In response to these limitations of mechanical and fixed-function flight computers  --- 
and building on recent findings that software-based performance calculation tools
can match approved flight manual data while reducing chart interpretation error
\cite{BabbHiers2018}  --- 
this paper presents E6BJA: a cross-platform, offline-capable digital software-based flight computer. Designed to integrate the pedagogical value of mechanical tools with the usability, extensibility, and cognitive scaffolding of modern software, E6BJA reimagines the flight computer as both a computational aid and an educational environment.

The E6BJA app is available for Apple iOS, Android, and Microsoft SurfacePro/Windows devices, and is structured around a growing suite of calculators tailored to the needs of GA pilots, instructors, and flight planning educators. These include legacy functions such as true airspeed and wind correction, as present in the electronic flight calculators, but also more advanced tools such as holding pattern calculator, wind and diversion visualiser, carburettor icing risk estimation, take-off and landing performance safety factors, as well as common GA aircraft-specific weight and balance profiles. E6BJA functions are also supported by an embedded educational system of explanatory monographs.\newline

This paper provides a comprehensive comparison between E6BJA and its mechanical and electronic predecessors. It evaluates the systems across functional, cognitive, and technical dimensions, identifying where E6BJA provides added value and where traditional tools remain appropriate. In doing so, we aim to demonstrate how a software-based approach to flight computers can improve safety, learning, and usability  ---  without abandoning the rigour and precision aviation demands.\newline

Furthermore, E6BJA offers tailored weight and balance calculators for over 25 popular GA aircraft types, including the C152, C172S, PA-28 variants, DA-42, and Cirrus SR22. This eliminates the need to reference multiple POH sheets or create external spreadsheets, enhancing safety and standardisation in club or flight school operations.

\section{Design Goals and Principles}

The design and development of the E6BJA Flight Computer App has been guided by a central aim: to create a modern, extensible flight computer that enhances the reliability, educational value, and usability of aeronautical computations, while preserving the conceptual rigour of traditional flight planning tools. This section outlines the core design principles that shaped the platform, from interface design and platform reach to functional architecture and pedagogical integration.

\subsection{Accuracy, Reliability, and Reduction of Ambiguity}

Aviation computations demand high precision, reproducibility, and clarity. Traditional mechanical flight computers, such as the CRP-1 and E6B, though robust, are vulnerable to misinterpretation  ---  especially when reading scales with similar numerical values or performing multi-step calculations under time pressure. E6BJA ensures accuracy by embedding fixed models with clearly defined units and decimal logic, removing ambiguity in both input and output. The calculators in E6BJA incorporate verified, authoritative models, including the 1976 International Standard Atmosphere and UK CAA guidelines for take-off and landing performance \cite{UKCAASafetySense09}. By grounding its computations in these standards, E6BJA delivers greater precision for critical parameters such as pressure and density altitude, carburettor icing risk, and runway performance margins  ---  far surpassing the rough estimations inherent in traditional whiz-wheels.

\subsection{Multi-Platform Native Implementation}

Portability and cross-device availability were essential to the E6BJA design. The application is implemented natively for Apple iOS, Android, and Microsoft Windows (Surface Pro) platforms \cite{E6BJAiOS,E6BJAAndroid,E6BJAWindows}, ensuring optimal performance and responsive interfaces across both tablets and smartphones. This native development approach avoids compromises associated with cross-platform toolkits and enables seamless adaptation to both touch-based and stylus/keyboard environments. All core architectures are supported, including x86, x64, ARM, and ARM64, reflecting the diverse hardware ecosystems used by modern pilots and student aviators.

\subsection{User-Centred Design: Accessibility, Discoverability, and Visual Clarity}

E6BJA adopts a user-centred interface model, designed to support both novice learners and experienced pilots. Large interactive elements, logically grouped toolsets, and minimal navigation depth ensure that key functions are discoverable without requiring extensive training or documentation. Unlike older dedicated digital tools, the app benefits from full use of colour, scalable vector graphics, and mobile-native UI components. Where applicable, visualisations  ---  such as in the Wind and Diversion Visualiser or Holding Pattern Computer   ---  aid understanding by making abstract concepts spatially tangible.\newline

User-centred design principles  ---  such as discoverability, feedback, and visual clarity  ---  are critical for ensuring flight training tools are not only operationally usable, but also pedagogically effective \cite{Schwartzentruber2017}. Recent work on remote pilot training platforms highlights that perceived ease of use and usefulness are among the strongest predictors of adoption by instructors and trainees \cite{WaghPackirisamy2025}. Moreover, research on EFB training practices at collegiate flight schools has shown that the presence of formal policies does not necessarily correlate with effective EFB usage or instructional quality. Instead, well-integrated training practices  ---  especially those that emphasise fundamental competencies  ---  are more predictive of successful adoption and safe use \cite{Babb2017}.

\subsection{Educational Support Through Embedded Monographs}

Each calculator in the E6BJA app is accompanied by a concise embedded monograph, providing users with explanatory content on the calculator’s function, assumptions, and underlying equations. These integrated texts are designed to enhance both self-directed learning and structured aviation education. Where possible, explanatory monographs and context-aware prompts guide user comprehension, especially for calculations that are typically opaque in other tools. This feature supports training, self-guided revision, and instruction. Crucially, it addresses a key concern in digital aviation tools: the risk of over-reliance on automation leading to degradation of manual cognitive skills, as demonstrated in empirical studies of flight planning performance \cite{Volz2016}. This dual role helps preserve cognitive engagement with fundamental concepts while reducing the risk of skill degradation often associated with over-automation.

\subsubsection{Aircraft-Specific Weight and Balance Modelling}
\label{sec:weight_balance}

Accurate weight and balance computation is fundamental to safe aircraft operation, yet it is
frequently under-supported by generic flight computers. Traditional mechanical and electronic
calculators typically require pilots to perform weight and balance calculations externally using
paper graphs or manually constructed spreadsheets derived from the Pilot’s Operating Handbook
(POH).

\begin{figure}[htbp]
\centering
\includegraphics[width=0.75\linewidth]{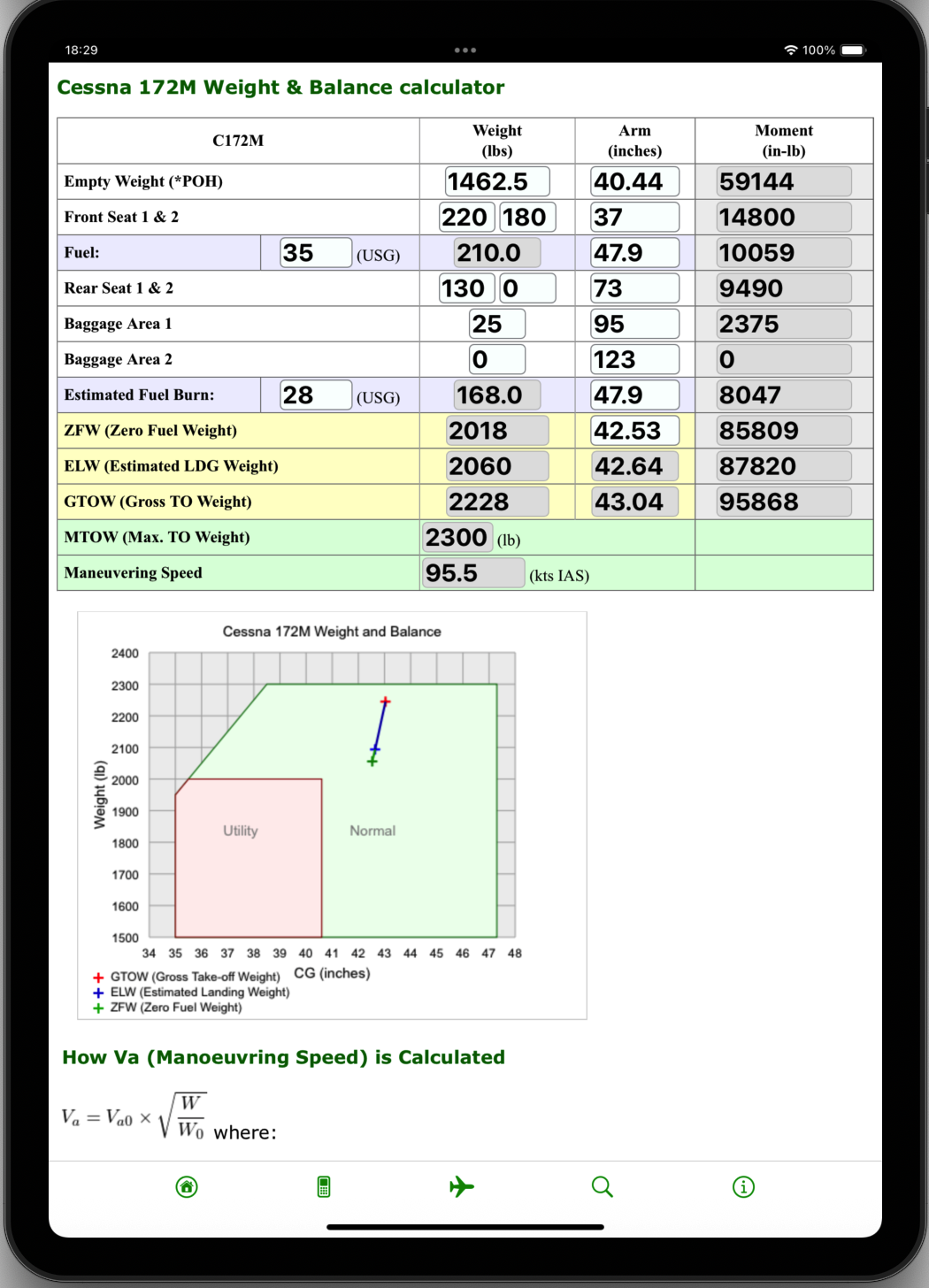}
\caption[Aircraft-specific weight and balance modelling in E6BJA]{\label{fig:weight_balance_c172m}
Aircraft-specific weight and balance calculation for a Cessna~172M in E6BJA. The calculator encodes
POH-derived mass limits and CG envelope geometry, presenting both numeric loading data and a
graphical centre-of-gravity plot. This approach supports rapid verification of loading legality
and improves understanding of CG movement under fuel burn and payload variation.}
\end{figure}

E6BJA incorporates aircraft-specific weight and balance models for a wide range of general aviation
aircraft, including the Cessna 172M, 172S, PA-28 variants, and other commonly operated training and
touring aircraft. Each model encodes the aircraft’s certified mass limits, arm definitions, and
centre-of-gravity (CG) envelope directly from POH data, allowing pilots to compute zero-fuel,
take-off, and landing conditions within a single integrated environment. Figure~\ref{fig:weight_balance_c172m} illustrates a representative aircraft-specific weight and balance calculation in E6BJA, showing both numeric outputs and graphical centre-of-gravity envelope visualisation for a Cessna~172M.

The calculator provides both numerical outputs and graphical CG envelope visualisation, enabling
users to assess not only whether a loading configuration is legal, but how close it lies to
certification boundaries. This graphical feedback supports conceptual understanding of longitudinal
stability, loading trade-offs, and fuel burn effects, which are difficult to convey using purely
numeric outputs.

\subsection{Carburettor Icing Risk Approximation}
\label{sec:carb_icing}

Carburettor icing remains a persistent hazard in light aircraft equipped with carburetted engines,
particularly during operations at reduced power settings. Unlike many aerodynamic or performance
phenomena, carburettor icing is governed by a combination of thermodynamic effects, airflow
characteristics, engine design, and operational context. As a result, no simple deterministic model
can fully predict its onset or severity.

E6BJA includes a \emph{Carburettor Icing Risk Approximator} designed explicitly as an educational and
situational-awareness tool rather than a predictive engine model. The calculator estimates the
likelihood of carburettor icing using ambient air temperature and dew point, from which relative
humidity and dew point spread are derived. These values are then mapped onto empirically established
risk regions commonly presented in aviation training literature and safety guidance.

\begin{figure}[htbp]
\centering
\includegraphics[width=0.75\linewidth]{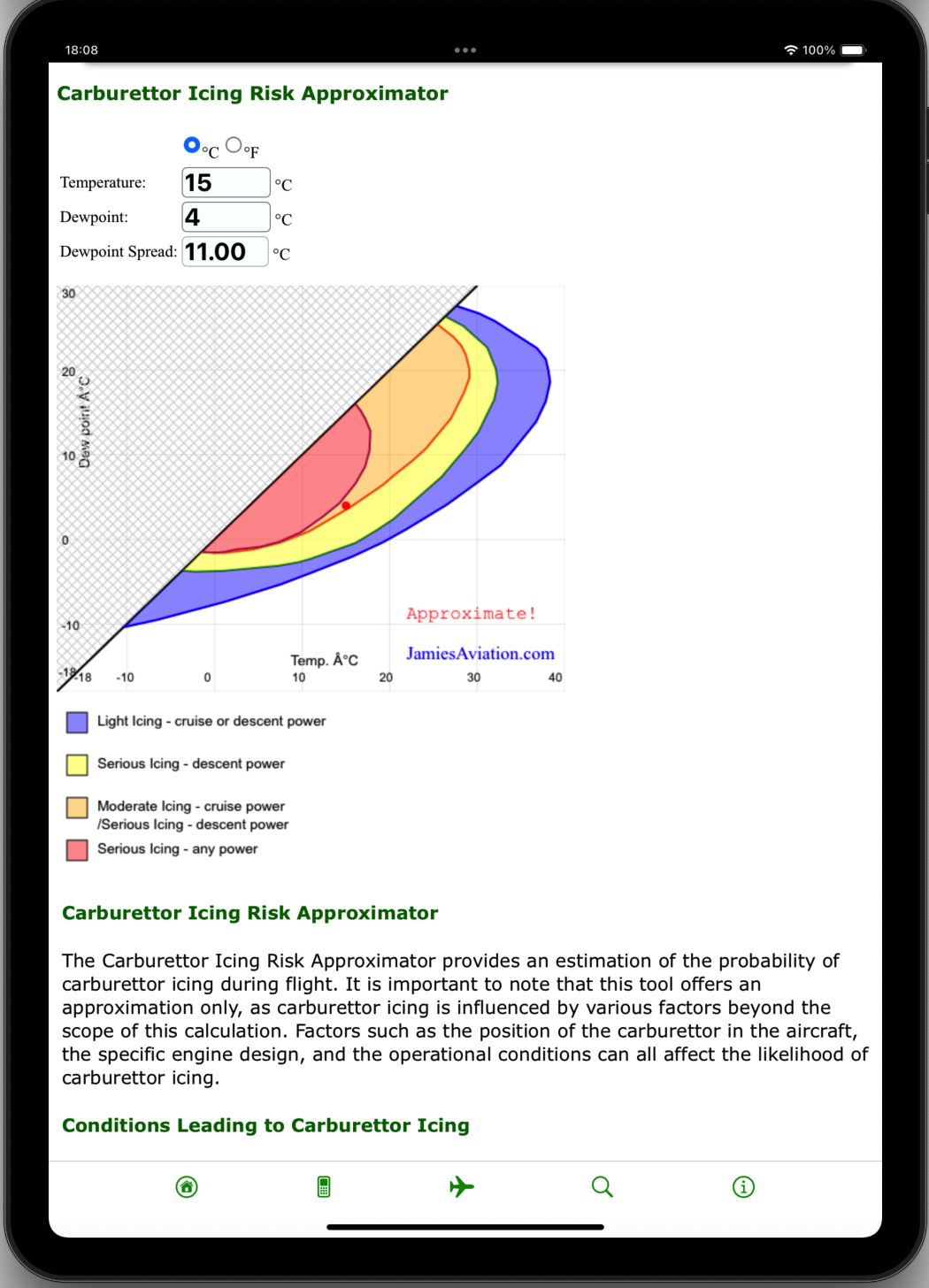}
\caption[Carburettor icing risk approximation in E6BJA]{\label{fig:carb_icing_risk}
Carburettor icing risk approximation in E6BJA. The tool maps ambient temperature and dew point onto
qualitative icing risk regions commonly used in flight training, illustrating the increased
likelihood of icing under high humidity and moderate temperature conditions. The output is
explicitly presented as an approximation and is intended to support situational awareness and
instruction rather than deterministic prediction.}
\end{figure}

The resulting output classifies icing risk into qualitative categories, including \emph{light icing},
\emph{moderate icing}, and \emph{serious icing}, with distinctions drawn between cruise-power and
descent-power operation. A boundary corresponding to 100\% relative humidity is also indicated,
representing conditions under which carburettor icing is considered imminent. An example output from the Carburettor Icing Risk Approximator is shown in Figure~\ref{fig:carb_icing_risk}, illustrating how temperature and moisture conditions are mapped onto established icing-risk regions.

Crucially, the tool is explicitly labelled as an approximation. Factors such as carburettor location
within the engine cowling, specific engine and induction system design, throttle setting, fuel
vaporisation characteristics, and pilot technique are not modelled. These limitations are disclosed
directly within the application to avoid false precision or over-reliance on numerical output.

By visualising risk as a function of temperature and moisture rather than presenting a binary
decision, the Carburettor Icing Risk Approximator reinforces correct operational thinking: that
carburettor icing is a probabilistic hazard requiring continuous monitoring and conservative
technique, not a condition that can be ruled in or out by calculation alone.

\subsection{Aircraft-Specific Configurability}

One of the major limitations of conventional digital flight computers is their inability to adapt to specific airframes. E6BJA addresses this through a growing library of aircraft-specific performance modules, each configured for the characteristics of a particular aircraft model. As of 2025, the app includes performance profiles for over 25 general aviation aircraft, including multiple variants from Cessna, Piper, Beechcraft, Cirrus, Diamond, Robin, and Socata. Each profile includes customised weight and balance tools, as well as performance calculators for take-off distance, climb rate, fuel consumption, and landing performance, reflecting manufacturer data and standard conditions. This enables more realistic and tailored planning during pilot training and transition exercises.

\subsection{Functional Architecture: Two Integrated Flight Computers}

E6BJA’s architecture is structured around two core domains:

\begin{itemize}
  \item The E6BJA Flight Computer
  \item The Flight Performance Computer
\end{itemize}

The Flight Computer offers over 40 modules covering a wide range of aeronautical and flight calculations. These include wind correction and triangle solvers, fuel and glide planning, density-altitude converters, Mach number computations, ISA models, dew point and humidity estimators, and various geometry and navigation calculators (e.g., rhumb line, great circle, line-of-sight). These tools are tightly integrated with explanatory content and consistent UI components, maintaining both usability and accuracy.\newline

The Flight Performance Computer focuses on aircraft-specific planning. Each aircraft module includes calculators tailored to that airframe’s geometry, weight limits, and performance charts, including take-off and landing distances, climb profiles, and cruise planning tools. For example, the Cessna 172S module includes distinct calculators for take-off roll (at 2,550 lbs), climb rate, and fuel usage during ascent. These calculators are useful both for preparing real-world performance plans and for teaching students how to interpret and apply flight manual data in context (TOLD App Study, 2016).
Together, these two domains position E6BJA not merely as a replacement for mechanical or digital flight computers, but as a modular educational platform for aeronautical computation. Its structure is designed to support progressive learning, from foundational navigation and atmospheric concepts through to airframe-specific planning exercises used in advanced ground school and flight instructor training.

\subsubsection{Performance Safety Factors and Real-World Margin Modelling}
\label{sec:caa_safety_factors}

UK Civil Aviation Authority Safety Sense Leaflet 09 (\emph{Aeroplane Performance}, superseding 7c) provides
explicit quantitative guidance on how pilots should adjust take-off and landing distances derived
from unfactored manufacturer data to account for real-world operating conditions and appropriate
safety margins \cite{UKCAASafetySense09}. The leaflet emphasises that performance-related accidents frequently
arise not from lack of data, but from failure to correctly apply multiple interacting correction
factors, which must be \emph{multiplied}, not summed, to obtain realistic Take-Off Distance Required
(TODR) and Landing Distance Required (LDR) values.

While Safety Sense Leaflet~7c includes tabulated factors and worked examples, its practical
application places a significant cognitive burden on pilots. Each contributing variable  ---  aircraft
weight, aerodrome elevation, ambient temperature, runway surface condition, slope, and tailwind
component  ---  introduces a multiplicative adjustment, and the combined effect is often underestimated
when performed manually or applied inconsistently. Traditional mechanical flight computers do not
support this process directly, and electronic flight calculators typically assume idealised
conditions or require pilots to apply safety margins externally.

E6BJA addresses this gap by explicitly encoding the CAA’s factor-based methodology as a structured
computational model. Rather than treating performance margins as informal annotations, each
contributing factor is implemented as an independent multiplicative term derived directly from
Safety Sense Leaflet~09 (previously 7c). These include proportional distance increases for aircraft weight,
incremental penalties for elevation and temperature, tailwind effects calculated as a function of
tailwind expressed as a percentage of lift-off speed ($V_{LO}$), asymmetric treatment of runway
slope, and discrete surface-condition multipliers for grass, wet surfaces, soft ground, or snow.

In accordance with CAA guidance, the calculator enforces a strictly multiplicative composition of
factors and explicitly presents the resulting factor chain (e.g.\ $\times 1.2 \times 1.1 \times
1.3$) to the user. The calculation is performed in two stages. First, an environmentally adjusted
distance is computed, reflecting performance degradation due to prevailing conditions. Second, the
CAA-recommended general safety factor is applied  ---  $\times 1.33$ for take-off and $\times 1.43$ for
landing  ---  yielding a conservative TODR or LDR suitable for pre-flight planning.

This separation mirrors the conceptual distinction in Safety Sense Leaflet~09 between environmental
performance penalties and regulatory safety margins applied to unfactored data. By maintaining this
distinction, E6BJA avoids conflating certification buffers with operational conditions, a common
error in ad-hoc spreadsheet or mental calculations.

By operationalising regulatory safety guidance in software, E6BJA reduces reliance on manual
arithmetic while preserving transparency of assumptions and margins. The calculator is not intended
to replace aircraft-specific performance schedules contained in the Pilot’s Operating Handbook or
Flight Manual, which remain authoritative. Instead, it provides a repeatable and auditable framework
for applying CAA-recommended safety factors when unfactored data are used, particularly in
educational, instructional, and general aviation planning contexts.

\begin{figure}[htbp]
\centering

\begin{subfigure}[t]{0.47\textwidth}
    \centering
    \includegraphics[width=\linewidth]{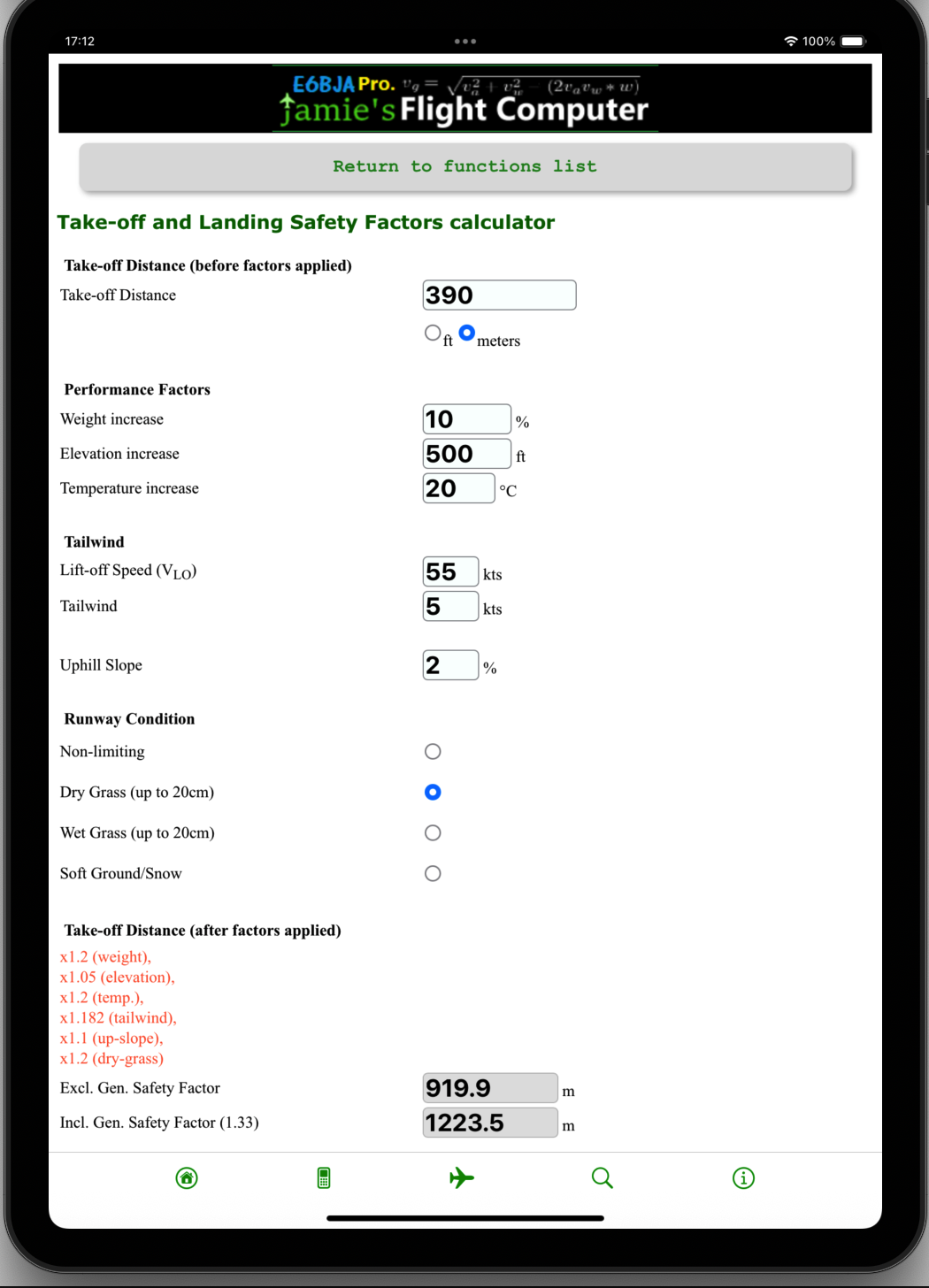}
    \caption{Take-off distance calculation in E6BJA using metric units (metres). The calculator applies CAA Safety Sense Leaflet~09 performance and safety factors multiplicatively, explicitly listing each contributing factor and producing both condition-adjusted and fully factored TODR values.}
    \label{fig:e6bja_takeoff_told_m}
\end{subfigure}
\hfill
\begin{subfigure}[t]{0.47\textwidth}
    \centering
    \includegraphics[width=\linewidth]{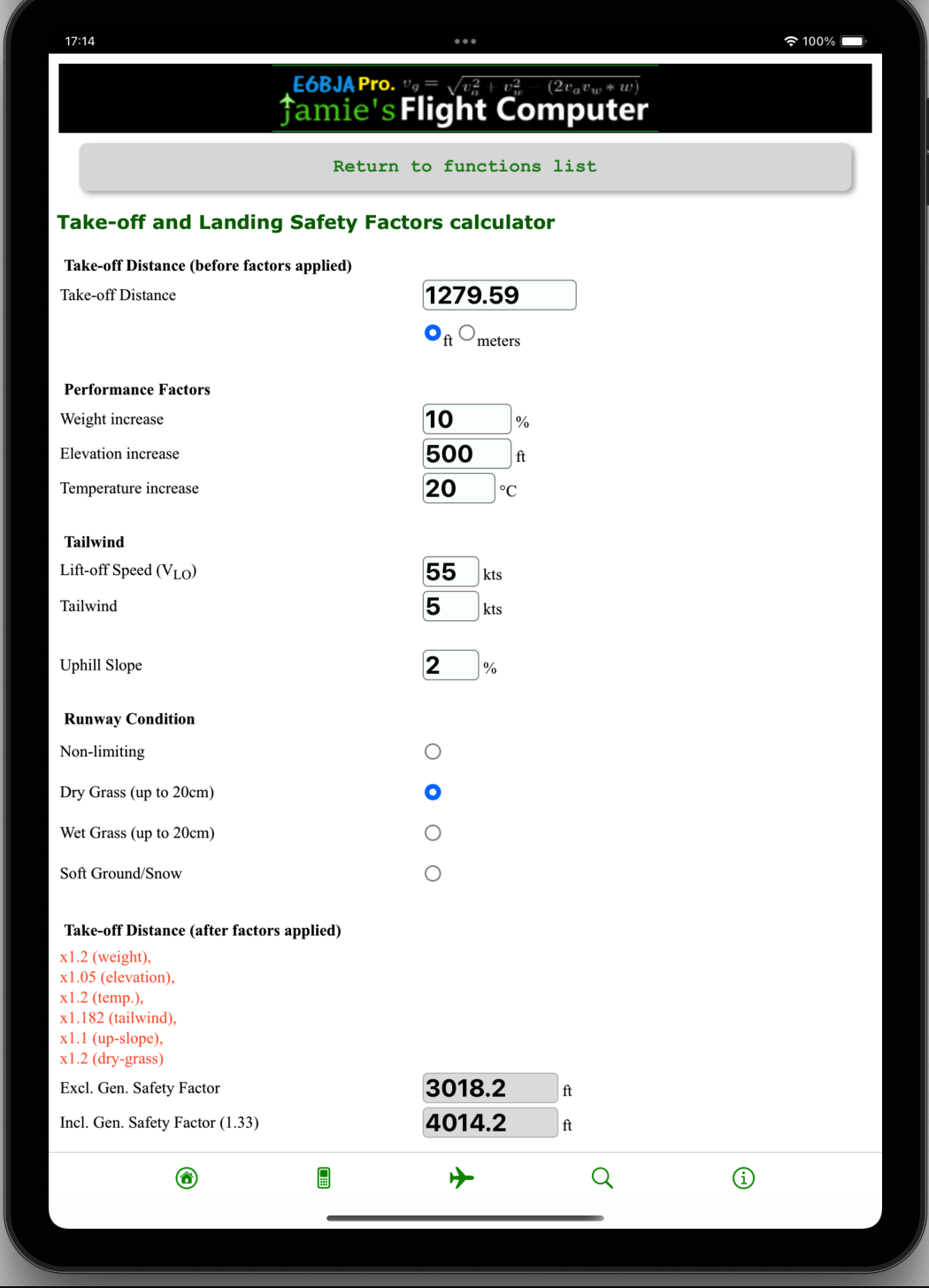}
    \caption{The same take-off scenario expressed in imperial units (feet). Unit conversion is performed transparently, yielding numerically consistent results while preserving the full factor chain and regulatory safety margins.}
    \label{fig:e6bja_takeoff_told_ft}
\end{subfigure}

\caption[Take-off distance calculation with safety factors in E6BJA]{\label{fig:e6bja_takeoff_told_comparison}
Comparison of take-off distance calculations in E6BJA using metric and imperial units for an identical operational scenario. Moderate increases in aircraft weight, elevation, temperature, tailwind, uphill slope, and grass runway surface are applied multiplicatively in accordance with UK CAA Safety Sense Leaflet~09, increasing the unfactored take-off distance of approximately 390~m (1,280~ft) to a CAA-compliant Take-Off Distance Required (TODR) exceeding 1,220~m (4,010~ft). The paired views demonstrate both the non-intuitive compounding of performance penalties and the ease of unit conversion within the software-based flight computer.}
\end{figure}

The practical impact of applying these multiplicative performance and safety factors is illustrated
in Figure~\ref{fig:e6bja_takeoff_told_comparison}. The paired metric and imperial views demonstrate
how moderate environmental and operational penalties compound to substantially increase the
TODR, and how transparent unit handling supports safer pre-flight
planning.

\begin{figure}[htbp]
\centering

\begin{subfigure}[t]{0.47\textwidth}
    \centering
    \includegraphics[width=\linewidth]{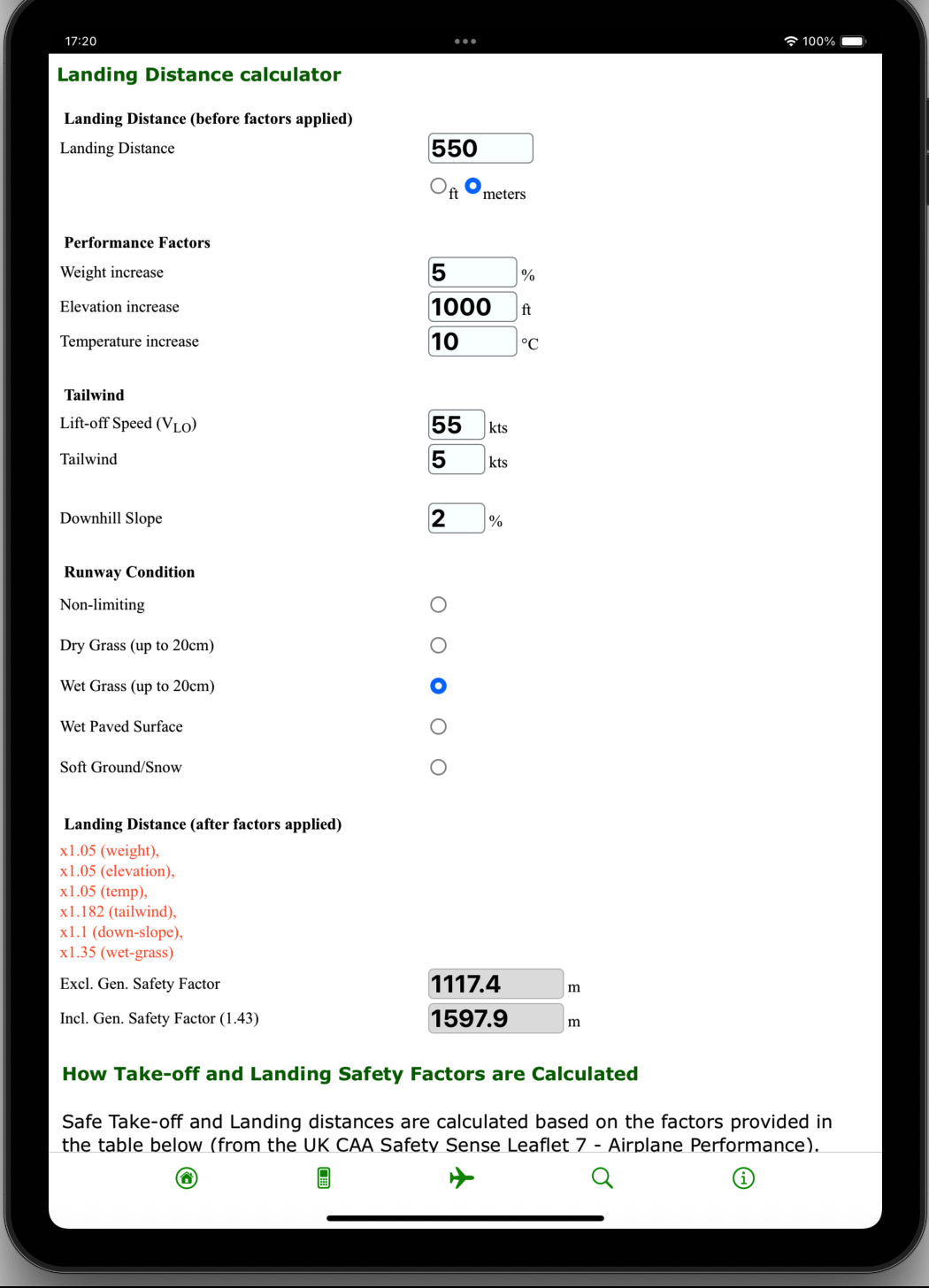}
    \caption{Landing distance calculation in E6BJA using metric units (metres). Performance and surface-condition factors are applied multiplicatively in accordance with CAA Safety Sense Leaflet~09, with separate reporting of the environmentally adjusted LDR and the fully factored, CAA-compliant landing distance.}
    \label{fig:e6bja_landing_told_m}
\end{subfigure}
\hfill
\begin{subfigure}[t]{0.47\textwidth}
    \centering
    \includegraphics[width=\linewidth]{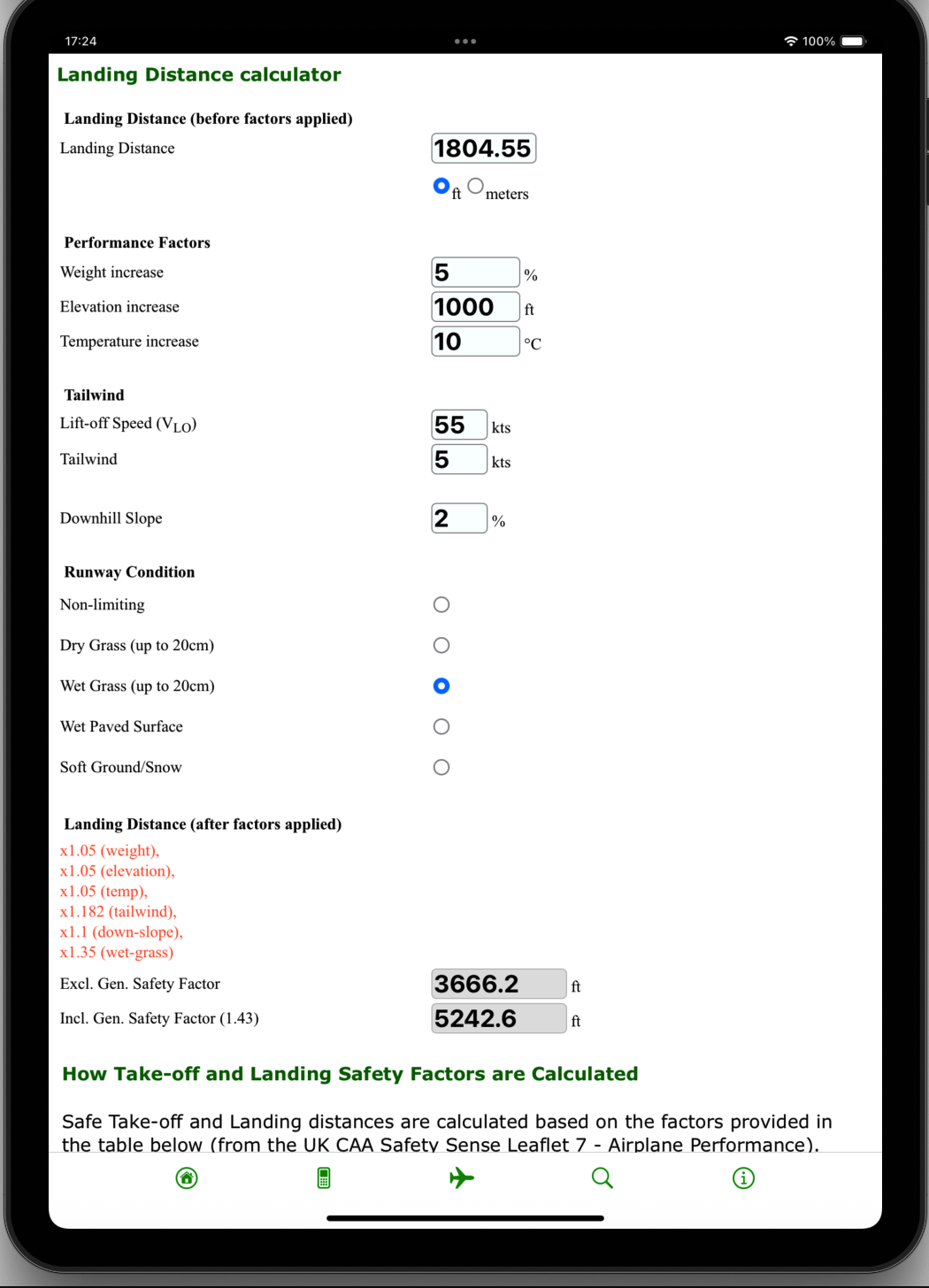}
    \caption{The same landing scenario expressed in imperial units (feet). Unit conversion preserves numerical consistency while retaining full visibility of the factor chain and the application of the CAA general landing safety factor.}
    \label{fig:e6bja_landing_told_ft}
\end{subfigure}

\caption[Landing distance calculation with safety factors in E6BJA]{\label{fig:e6bja_landing_told_comparison}
Comparison of landing distance calculations in E6BJA using metric and imperial units for an identical operational scenario. Moderate increases in aircraft weight, elevation, temperature, tailwind, downhill slope, and wet grass runway conditions are applied multiplicatively in accordance with UK CAA Safety Sense Leaflet~09. The unfactored landing distance of approximately 550~m (1,805~ft) increases to a CAA-compliant Landing Distance Required (LDR) approaching 1,600~m (5,240~ft), illustrating that landing performance is frequently more restrictive than take-off when regulatory safety margins are applied.}
\end{figure}

A corresponding landing-distance scenario is shown in
Figure~\ref{fig:e6bja_landing_told_comparison}. This example illustrates that, once regulatory safety
margins are applied, landing performance frequently becomes more restrictive than take-off
performance  ---  particularly on grass or contaminated runways  ---  reinforcing the importance of conservative
planning.

\subsubsection{Numerical Precision in the Application of CAA Performance Factors}

Safety Sense Leaflet~09 (previously 7c) specifies that take-off and landing performance penalties must be applied
\emph{multiplicatively} rather than additively. While the underlying model is unambiguous, its
manual application on paper or using tabulated references typically requires pilots to round or
approximate individual correction factors to the nearest recommended increment. This introduces
quantisation error that compounds across multiple factors.

\paragraph{Manual (paper-based) application.}
In a paper-based workflow, the Take-Off Distance Required (TODR) or Landing Distance Required (LDR)
is commonly estimated as
\begin{equation}
D_{\text{manual}} = D_0 \times \prod_{i=1}^{n} F_i^{\text{step}},
\end{equation}
where $D_0$ is the unfactored distance from the POH or performance chart, and each
$F_i^{\text{step}}$ is a discretised factor selected from tabulated guidance (e.g.\ rounding a
5--10\% tailwind effect to a fixed $\times 1.2$ multiplier).

\paragraph{Continuous software-based application.}
In contrast, E6BJA implements the same CAA model using continuous evaluation of each factor,
yielding
\begin{equation}
D_{\text{E6BJA}} = D_0 \times \prod_{i=1}^{n} F_i^{\text{cont}}(x_i),
\end{equation}
where $F_i^{\text{cont}}(x_i)$ is a continuous function of the underlying variable $x_i$
(e.g.\ tailwind expressed as a proportion of lift-off speed $V_{LO}$).

For example, the CAA tailwind correction is defined as a $20\%$ increase in distance for a tailwind
equal to $10\%$ of $V_{LO}$. In E6BJA, this is implemented as
\begin{equation}
F_{\text{tailwind}} = 1 + 0.2 \times \frac{V_{\text{TW}}}{0.1\,V_{LO}},
\end{equation}
allowing intermediate values (e.g.\ a $5~\mathrm{kt}$ tailwind with $V_{LO}=55~\mathrm{kt}$) to be
evaluated precisely as $\times 1.182$, rather than rounded to a coarser step value.

\paragraph{Implications.}
Both approaches conform to the CAA methodology. However, the continuous formulation reduces
rounding error and preserves the full multiplicative structure of interacting penalties,
particularly when several moderate factors are present simultaneously. This distinction becomes
material in safety-critical planning scenarios, where small differences at each stage can
compound into substantial changes in the final TODR or LDR.

\subsection{From Mechanical to Software-Based Landing Pattern Computation}

The Landing Pattern Computer applies the software-based design principles established in E6BJA to a distinct flight-planning domain, implemented as a native, multi-platform software application \cite{JALPCiOS,JALPCAndroid}.  It applies these principles specifically to the domain of teaching circuit planning and aerodrome joining procedures  ---  tasks that have traditionally been taught by training-aids such as the mechanical LPC-1 landing pattern computer. While the LPC-1 provides a fixed geometric depiction of standard left- or right-hand circuits relative to runway heading, it requires pilots to mentally integrate aircraft heading, wind conditions, and procedural sequencing.

The software-based implementation replaces this static representation with explicit geometric visualisation of circuit legs, wind vectors, and entry sectors, alongside procedurally sequenced guidance and wind-corrected headings for each leg of the circuit.  By making both track and heading information explicit, the system reduces reliance on mental reconstruction of circuit geometry and mitigates ambiguity at uncontrolled aerodromes — particularly under non-zero wind conditions or when multiple join options are available.

Figure~\ref{fig:lpc_mech_vs_software} contrasts the mechanical LPC-1 with a software-based landing pattern computer using identical runway and circuit parameters.

\begin{table}[htbp]
\centering
\begin{tabular}{cc}
\textbf{Mechanical Landing Pattern Computer (LPC-1)} &
\textbf{Software-Based Landing Pattern Computer} \\[0.5em]

\includegraphics[width=0.45\textwidth]{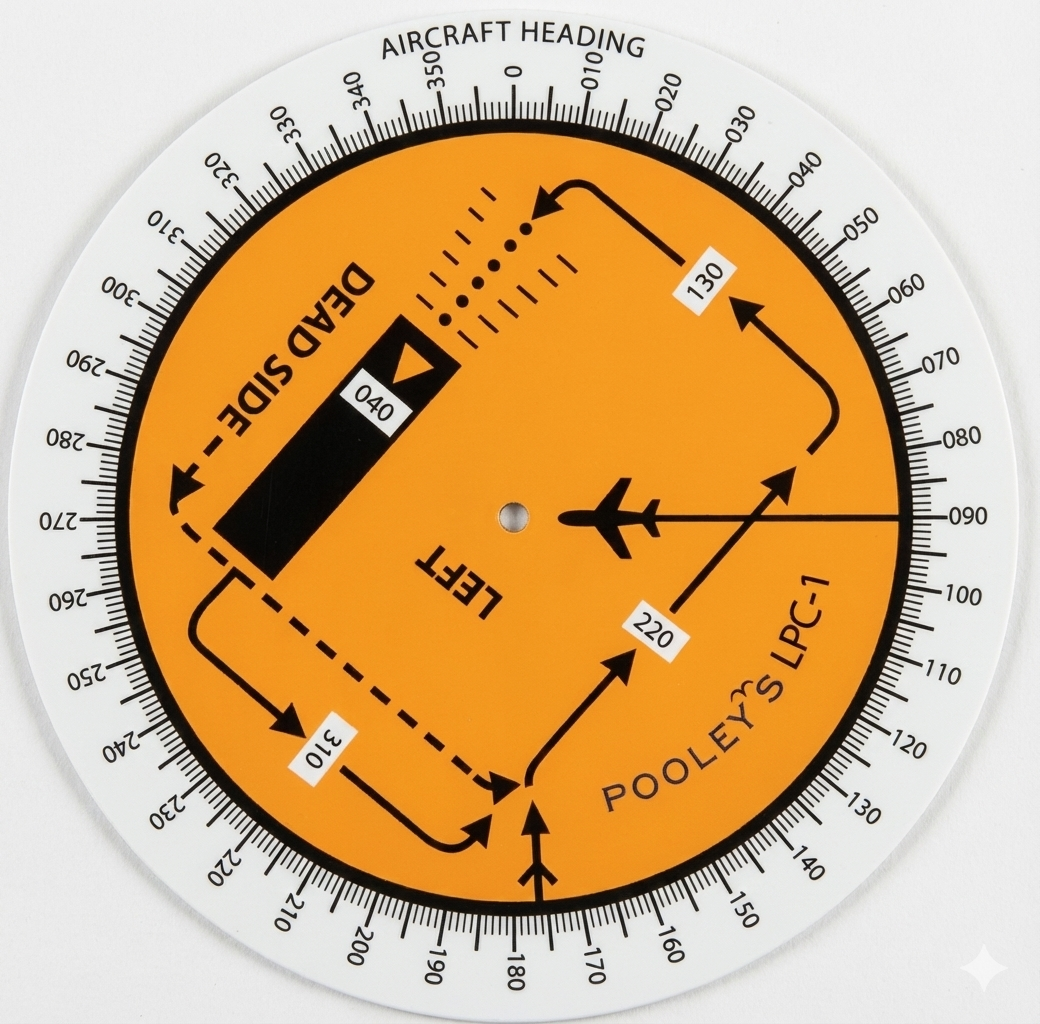} &
\includegraphics[width=0.45\textwidth]{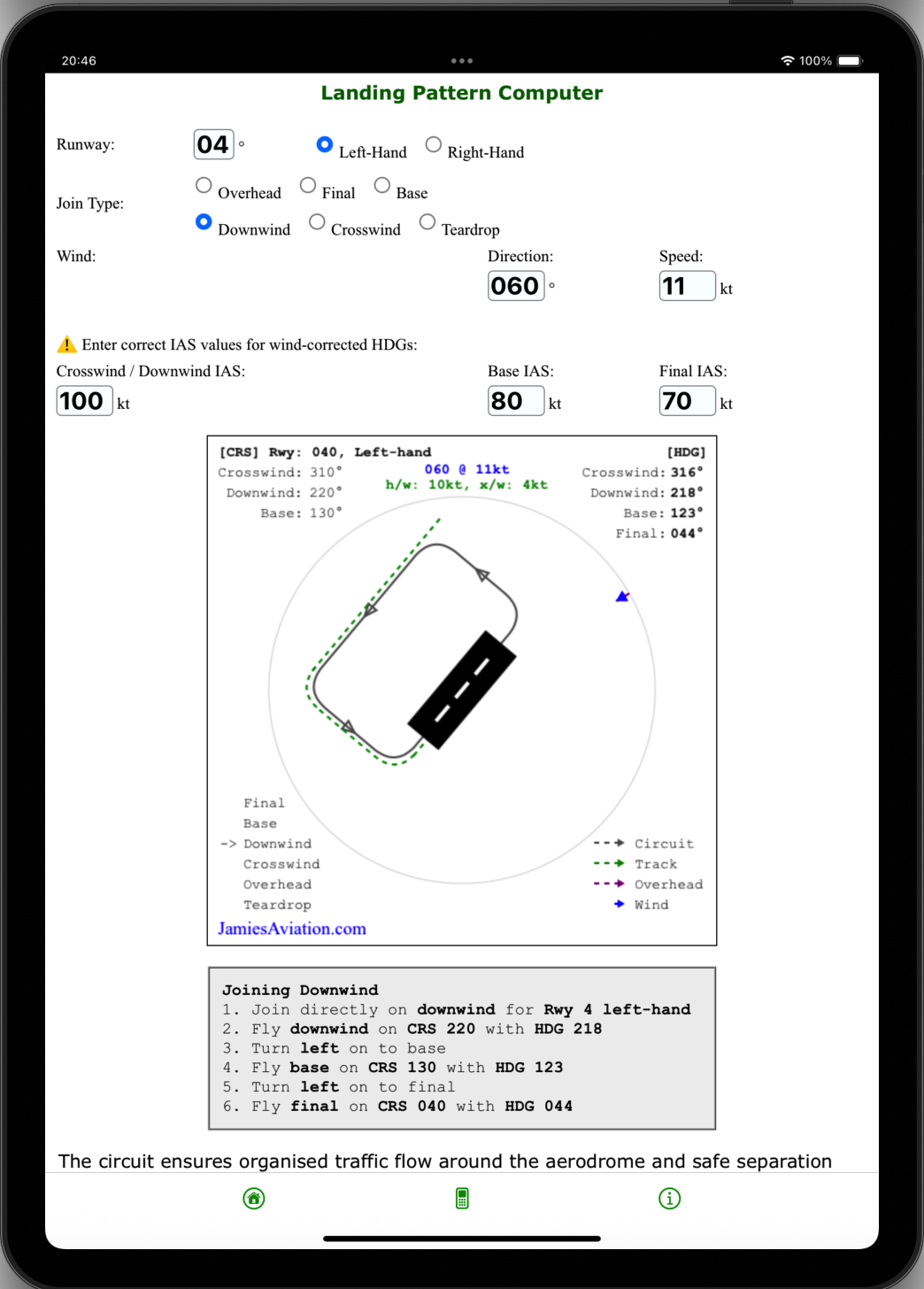} \\

\end{tabular}
\caption[Mechanical and software-based landing pattern computation]{\label{tab:lpc_comparison}
Comparison of landing pattern computation for Runway~040 with a left-hand circuit. 
\emph{Left:} The LPC-1 mechanical landing pattern computer depicts a fixed left-hand circuit geometry and derives circuit leg \emph{tracks} (crosswind~310$^\circ$, downwind~220$^\circ$, base~130$^\circ$) relative to runway heading, requiring pilots to infer headings and wind effects mentally.
\emph{Right:} The software-based Landing Pattern Computer reproduces the same circuit geometry and track structure while additionally modelling wind vectors, computing wind-corrected headings for each leg, and providing explicit, step-by-step procedural guidance for the selected join (downwind join shown).}
\end{table}

\begin{figure}[htbp]
\centering

\begin{subfigure}[t]{0.31\textwidth}
    \centering
    \includegraphics[width=\linewidth]{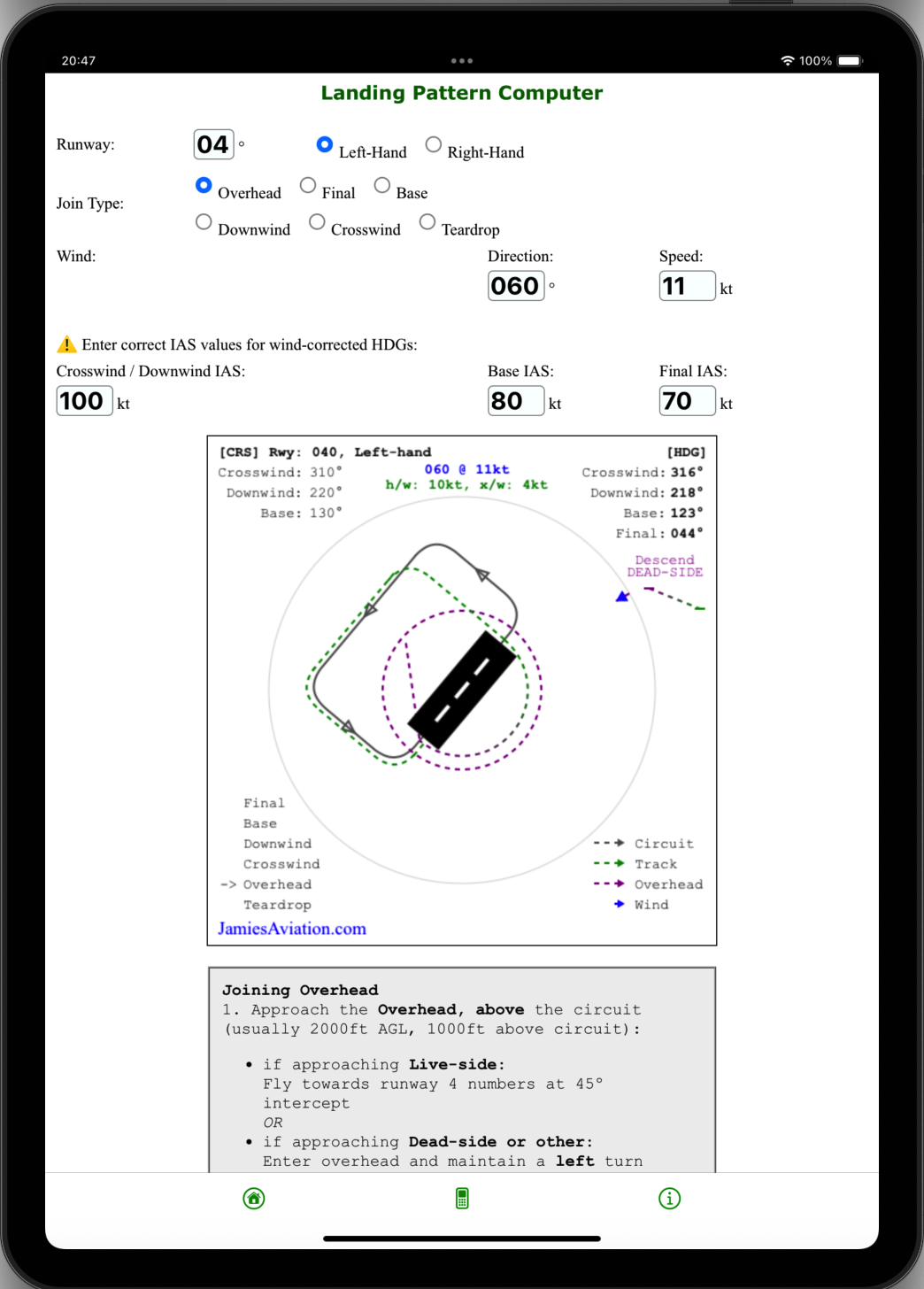}
    \caption{Overhead join visualisation. The software-based Landing Pattern Computer explicitly depicts entry geometry, circuit alignment, and procedural sequencing for overhead joins.}
    \label{fig:lpc_overhead}
\end{subfigure}
\hfill
\begin{subfigure}[t]{0.31\textwidth}
    \centering
    \includegraphics[width=\linewidth]{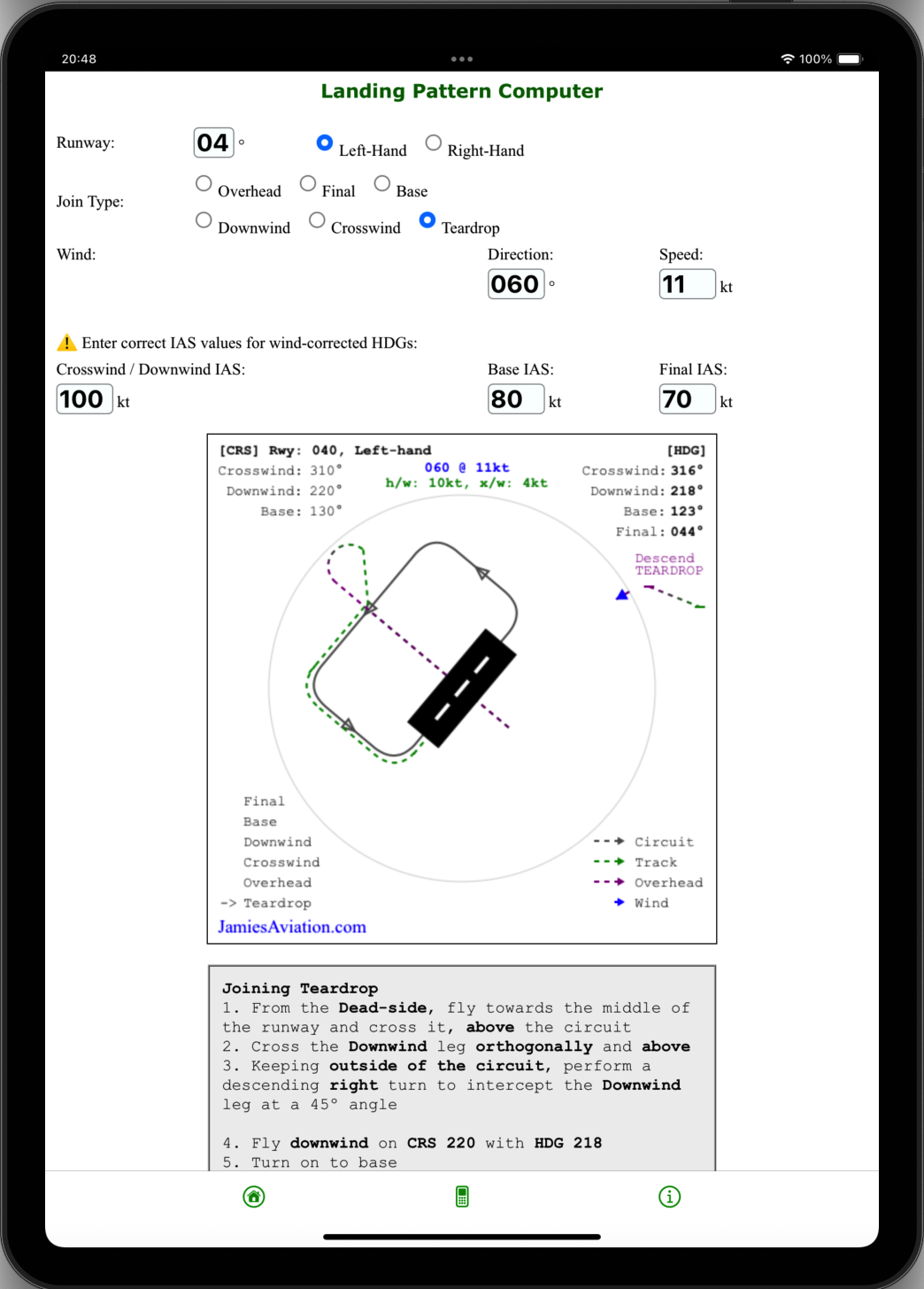}
    \caption{Teardrop join visualisation. Entry sectors and intercept geometry are shown explicitly, reducing ambiguity in join execution.}
    \label{fig:lpc_teardrop}
\end{subfigure}
\hfill
\begin{subfigure}[t]{0.31\textwidth}
    \centering
    \includegraphics[width=\linewidth]{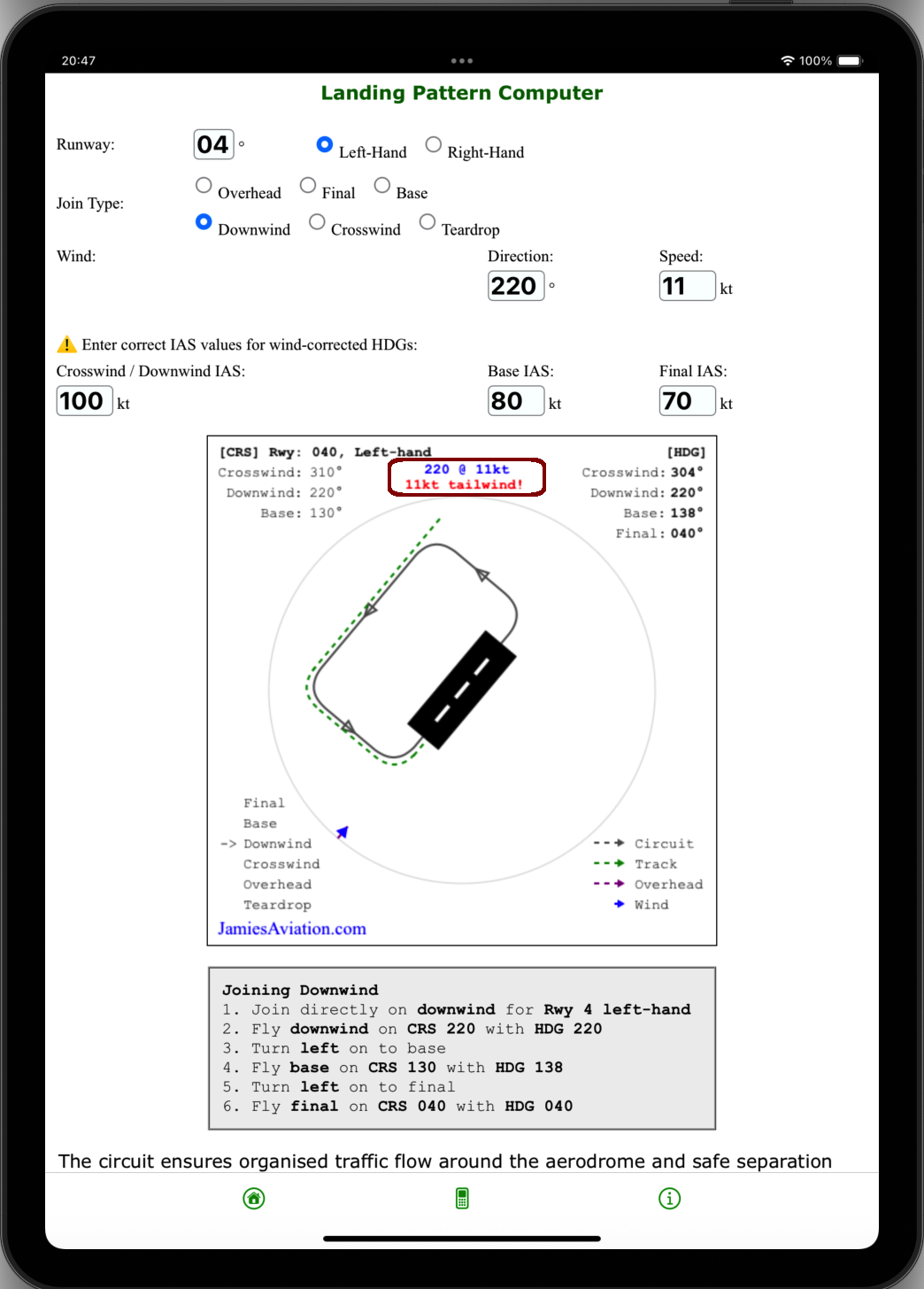}
    \caption{Downwind join with tailwind present. Wind vector, wind-corrected headings, and an explicit tailwind warning are displayed, supporting real-time situational awareness and safer decision-making.}
    \label{fig:lpc_tailwind}
\end{subfigure}

Figure~\ref{fig:lpc_software_extensions} illustrates software-only capabilities that extend beyond mechanical landing pattern computers, including join-specific geometry, wind-aware guidance, and explicit safety signalling.

\caption[Software-only extensions to landing pattern computation]{\label{fig:lpc_software_extensions}
Software-only extensions provided by the Landing Pattern Computer. Unlike mechanical devices, the software implementation supports explicit visualisation of multiple join procedures, wind-aware geometry, wind-corrected headings, and safety signalling (e.g.\ tailwind warnings). These features reduce cognitive load and mitigate ambiguity during circuit planning at uncontrolled aerodromes.}
\end{figure}

Together with the E6BJA Flight Computer, the Landing Pattern Computer illustrates how software-based aviation tools can preserve established procedural structures while materially improving clarity, error resistance, and instructional value.

\subsection{Architecture of a Software-Based Flight Computer}

The architecture of E6BJA is designed to exploit the processing power, flexibility, and sensor capabilities of modern mobile devices while preserving an offline-first design philosophy that prioritises reliability and access in austere environments. The system is implemented natively for three major platforms: Apple iOS (Swift), Android (Java and Android SDK), and Microsoft Windows (Universal Windows Platform, UWP)\cite{E6BJAiOS,E6BJAAndroid,E6BJAWindows}, with a complementary web-based version provided online.\newline

At the core of the system lies a shared JavaScript core aeronautical calculation engine, designed to be platform-agnostic and computationally rigorous. This engine ensures consistency of results across platforms while enabling rapid updates and the addition of new models without duplicating logic. Platform-specific interfaces wrap this core engine, allowing the app to conform to native UI and UX conventions on each device class, optimising responsiveness and user experience for tablets, phones, and hybrid touchscreen-keyboard systems such as the Surface Pro.\newline

Unlike fixed-function calculators, E6BJA’s modular software design enables layered functionality: core calculations (e.g., TAS, fuel burn, wind correction) are augmented by advanced modules (e.g., carburettor icing risk estimation, ISA-based altitude models, aircraft-specific weight and balance profiles), each of which integrates with a growing library of embedded monographs. Visualisation components  ---  including Holding Pattern Geometry, CG envelope plots, and Wind/Diversion overlays  ---  are powered by native rendering systems, ensuring high-resolution performance even on modest hardware.
The architecture also supports real-time validation, cross-field dependency checking, and assumption disclosure, helping users avoid common pitfalls such as inconsistent units or unmodelled conditions. Although all functionality is accessible offline, the system is engineered to allow future optional extensions, including cloud synchronisation, aircraft profile sharing, or integration with Electronic Flight Bags (EFBs), weather APIs, and digital logbooks.
This architecture reflects the dual nature of E6BJA: as both a calculation environment and an instructional system. Its design choices deliberately foreground clarity, performance, and extensibility, distinguishing it from legacy electronic calculators and demonstrating how modern mobile architectures can meaningfully improve flight planning, education, and pre-flight decision support.\newline

E6BJA leverages the substantial client-side processing power available on modern mobile devices, enabling it to handle complex aerodynamic models, real-time vector visualisations, and data validation without relying on server-side resources. This architecture not only ensures responsiveness and reliability in offline scenarios (such as during flight), but also allows the app to scale in complexity without compromising user experience. By exploiting native hardware acceleration and mobile OS SDK frameworks, E6BJA achieves both computational efficiency and extensibility  ---  capabilities far beyond those of dedicated electronic calculators.

\section{Comparative Evaluation}

\begin{figure}[htbp]
\centering

\begin{subfigure}[t]{0.45\textwidth}
    \centering
    \includegraphics[width=\linewidth]{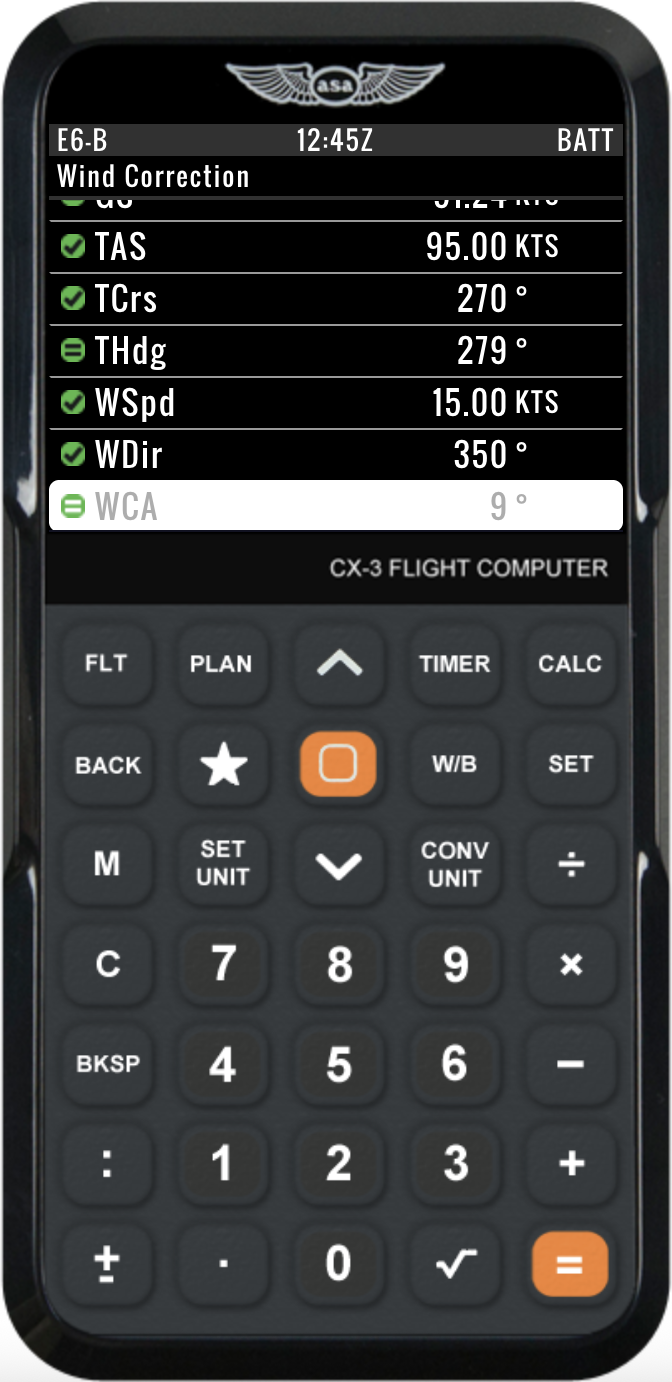}
    \caption{Wind correction angle (WCA) calculated by a traditional electronic flight calculator (ASA CX-3). The device reports numeric wind correction and ground speed values only, without visualisation of wind vectors or spatial context.}
    \label{fig:cx3_wca}
\end{subfigure}
\hfill
\begin{subfigure}[t]{0.45\textwidth}
    \centering
    \includegraphics[width=\linewidth]{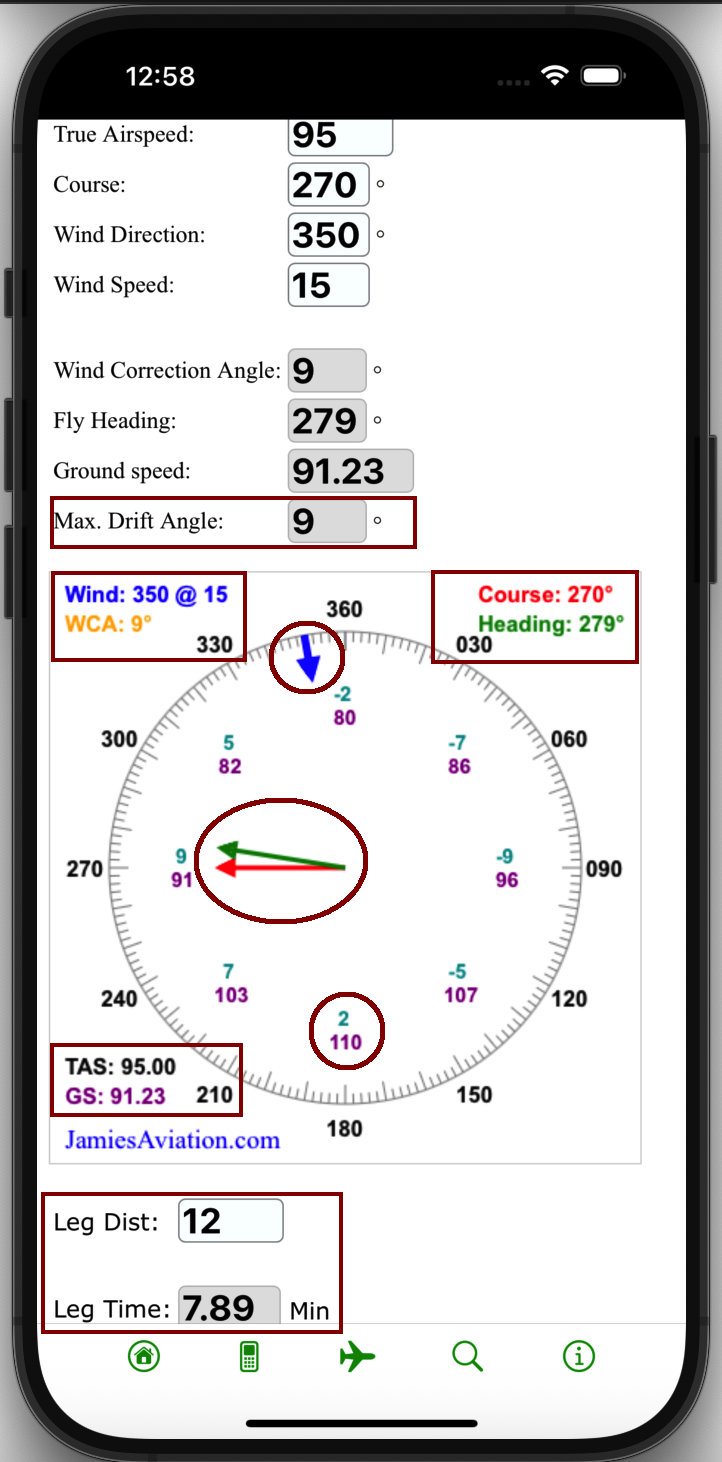}
\caption{Wind correction angle solution in E6BJA using the E6B Diversion Star. Numeric outputs equivalent to the ASA CX-3 (WCA, heading, and ground speed) are supplemented by a geometric wind-triangle visualisation. Pointer needles indicate intended course and corrected heading, while wind direction and magnitude are shown explicitly. The diversion star encodes WCA and ground speed for all bearings around the compass rose, reports maximum drift angle for mental approximation, and supports leg-distance input to derive time en route.}
    \label{fig:e6bja_wca_diversion_star}
\end{subfigure}

\caption[Comparison of wind correction angle presentation]{\label{fig:wca_comparison}
Comparison of wind correction angle (WCA) calculation for an identical navigation scenario (TAS~95~kt, course~270$^\circ$, wind~350$^\circ$ at~15~kt). The electronic flight calculator reports numeric WCA and ground speed values only. E6BJA presents the same solution using a diversion star that integrates numeric outputs with an explicit wind-triangle construction, visualising wind direction, corrected heading, drift magnitude, and ground speed across the compass rose. This supports spatial reasoning, rapid error detection, and mental navigation techniques commonly taught in training.}

\end{figure}

While electronic flight calculators collapse the wind triangle into numeric outputs, the
E6B Diversion Star in E6BJA preserves the geometric reasoning central to traditional
whiz-wheel navigation while extending it with continuous visual feedback. (Figure~\ref{fig:wca_comparison}) This design choice supports spatial reasoning, facilitates error detection, and aligns with the way wind correction is traditionally taught in flight training.

\begin{figure}[htbp]
\centering

\begin{subfigure}[t]{0.47\textwidth}
    \centering
    \includegraphics[width=\linewidth]{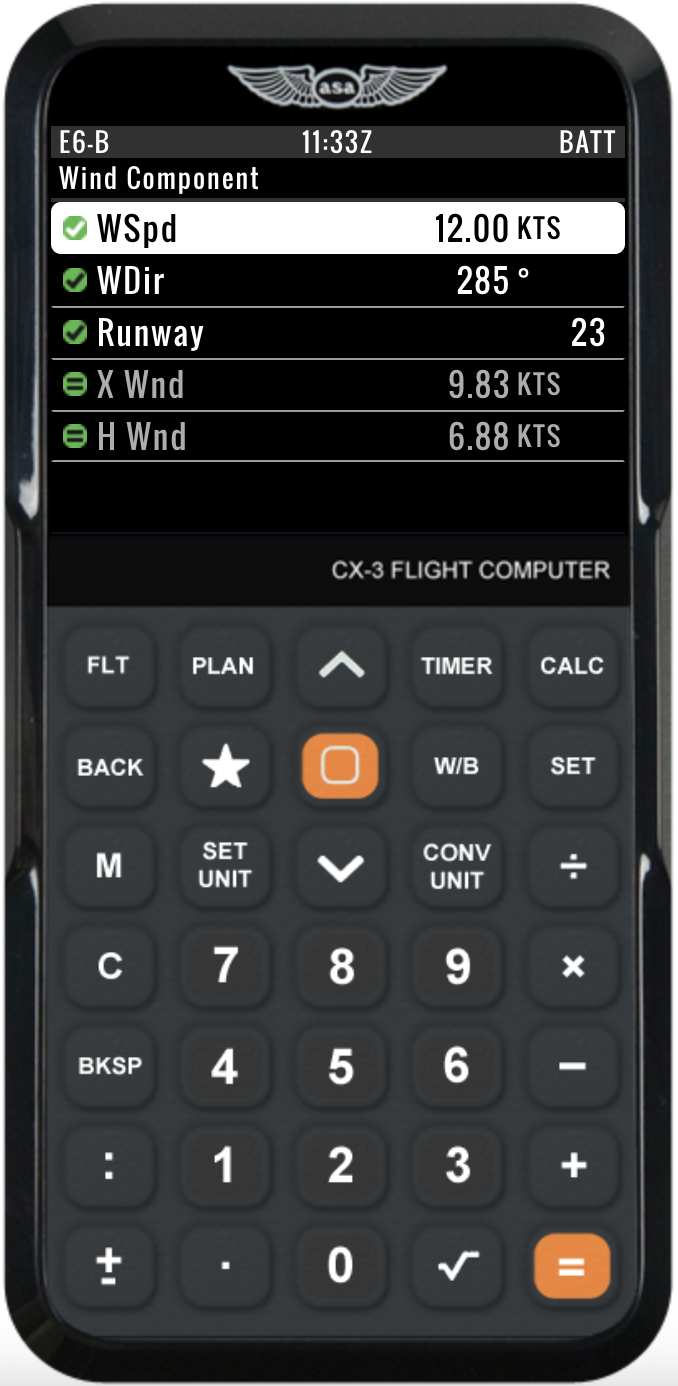}
    \caption{Wind components calculated by a traditional electronic flight calculator (ASA CX-3). The device reports numeric headwind and crosswind components only, without visual context or explanatory support.}
    \label{fig:cx3_wind_components}
\end{subfigure}
\hfill
\begin{subfigure}[t]{0.47\textwidth}
    \centering
    \includegraphics[width=\linewidth]{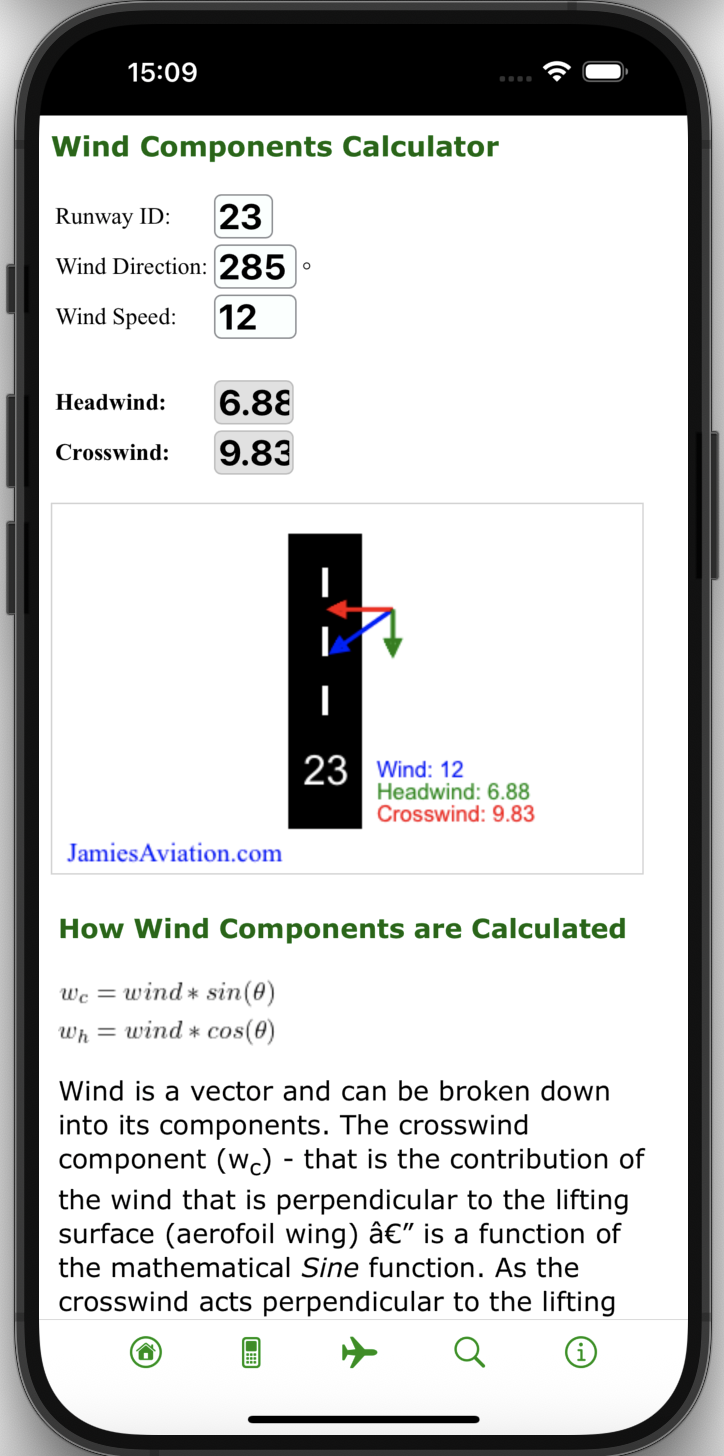}
    \caption{Wind components in E6BJA, showing numeric values alongside a graphical visualisation of the wind vector relative to the runway centreline, with embedded explanatory support.}
    \label{fig:e6bja_wind_components}
\end{subfigure}

\caption[Comparison of wind component presentation]{\label{fig:wind_components_comparison}
Comparison of wind component calculation and presentation for a representative scenario (Runway~23, wind~285$^\circ$ at~12~kt). The electronic flight calculator reports computed headwind and crosswind components numerically, whereas E6BJA supplements these values with an explicit vector visualisation aligned to the runway geometry. In addition, E6BJA provides an embedded monograph explaining the trigonometric basis of the calculation (sine and cosine decomposition) and the commonly taught “clock-code” mental approximation used by pilots for rapid crosswind estimation.}
\end{figure}

Differences in how wind components are presented are illustrated in Figure~\ref{fig:wind_components_comparison}. While traditional electronic flight calculators report numeric headwind and crosswind values only, E6BJA supplements these outputs with a geometric visualisation aligned to runway orientation.

Figure~\ref{fig:wca_comparison} contrasts the wind correction angle output of a traditional electronic calculator (Figure~\ref{fig:cx3_wca}) with the E6BJA diversion star visualisation (Figure~\ref{fig:e6bja_wca_diversion_star}), which preserves the geometric structure of the wind triangle while adding continuous visual feedback.

\begin{figure}[htbp]
\centering

\begin{subfigure}[t]{0.31\textwidth}
    \centering
    \includegraphics[width=\linewidth]{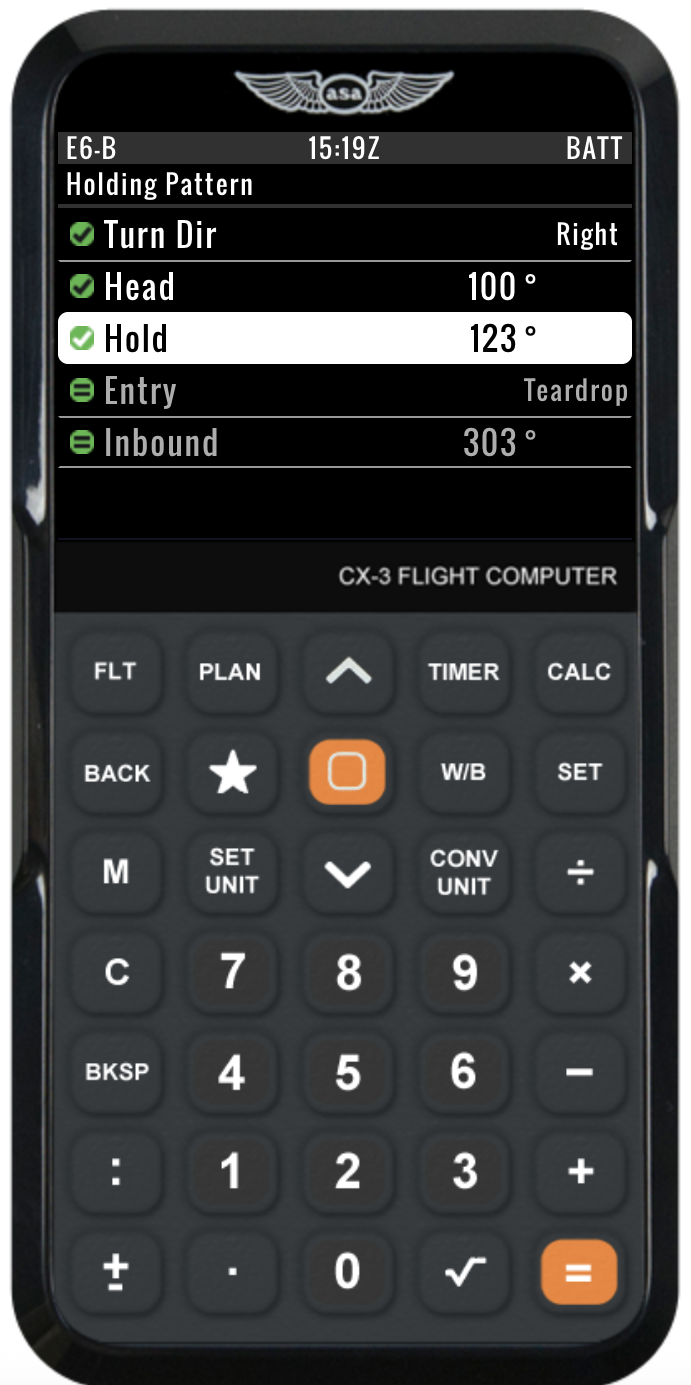}
    \caption{Electronic flight calculator output (ASA CX-3). Provides entry classification and inbound course only, with no wind correction, geometry, or procedural guidance.}
    \label{fig:cx3_hold}
\end{subfigure}
\hfill
\begin{subfigure}[t]{0.31\textwidth}
    \centering
    \includegraphics[width=\linewidth]{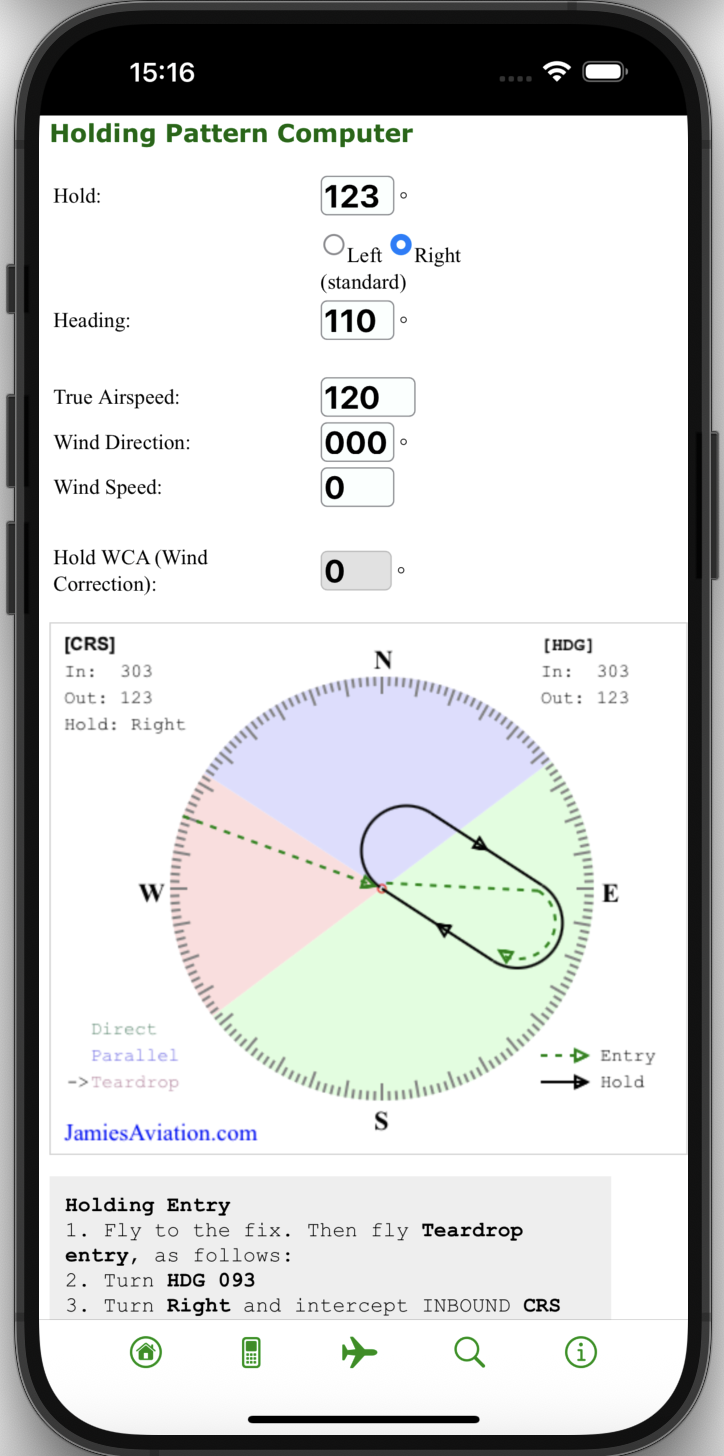}
    \caption{E6BJA holding pattern with zero wind (000$^\circ$ at 0~kt), used to ensure parity with the ASA CX-3. Geometry and procedural steps are visualised, but no wind correction is applied.}
    \label{fig:e6bja_hold_nowind}
\end{subfigure}
\hfill
\begin{subfigure}[t]{0.31\textwidth}
    \centering
    \includegraphics[width=\linewidth]{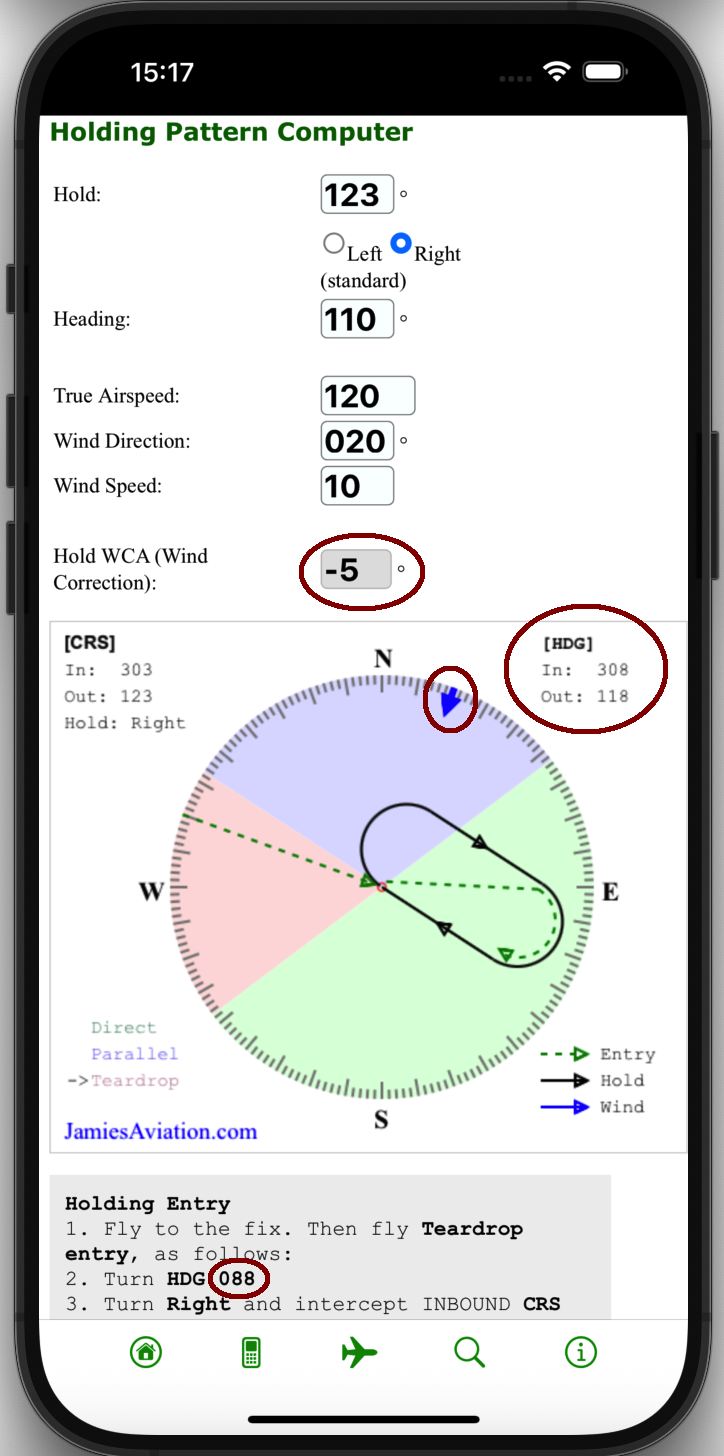}
    \caption{E6BJA holding pattern with wind applied (020$^\circ$ at 10~kt). Wind vector, wind correction angle, and wind-adjusted headings are explicitly visualised and reflected in procedural guidance.}
    \label{fig:e6bja_hold_wind}
\end{subfigure}

\caption[Holding pattern comparison with and without wind]{\label{fig:holding_pattern_threeway}
Comparison of holding pattern computation for a right-hand hold (heading~110$^\circ$, holding course~123$^\circ$, inbound course~303$^\circ$, teardrop entry). The ASA CX-3 provides a purely numeric classification. E6BJA reproduces this behaviour under zero-wind conditions for parity, and further extends it by modelling wind effects, visualising holding geometry, and generating wind-corrected procedural instructions.}
\end{figure}

Holding pattern computation highlights further differences between numeric and visual approaches. Figure~\ref{fig:holding_pattern_threeway} compares the ASA CX-3 output with E6BJA under both zero-wind and wind-corrected conditions.

\subsection{Comparison Table}

\begin{table}[htbp]
\centering
\makebox[\textwidth][c]{%
\scalebox{0.67}{%
\begin{tabular}{p{4.2cm}p{6.8cm}p{6.8cm}p{6.8cm}}
\hline
\textbf{Dimension} &
\textbf{Mechanical whiz-wheel (e.g., CRP-1 / E6B)} &
\textbf{Electronic flight calculators (Sporty’s Electronic E6-B, ASA CX-3)} &
\textbf{E6BJA (software-based flight computer)} \\
\hline

\textbf{Core computation model} &
Manual analogue slide-rule and wind grid; visual interpolation and physical vector construction; accuracy depends on scale resolution and user dexterity. &
Digital, menu-driven calculators with fixed computational routines; numeric input/output reduces interpolation ambiguity and speeds calculation. &
Software-defined computational models with explicit units and assumptions; supports higher-level modelling (e.g.\ ISA 1976, carburettor icing, aircraft performance modules) with integrated visual explanation. \\
\hline

\textbf{Wind triangle / WCA} &
Manual wind vector placement; errors arise from misplacement, incorrect scale selection (low/high speed), and angular misreading under time pressure or turbulence. &
Built-in wind/heading/groundspeed functions reduce construction error but remain sensitive to input accuracy and mode selection. &
Full wind triangle solver with graphical wind and groundspeed visualisation, enabling spatial interpretation of wind correction, drift, and resultant track. \\
\hline

\textbf{Functional breadth} &
Strong coverage of core E6B calculations (time--speed--distance, unit conversions, basic wind problems), bounded by printed scales and physical encoding. &
Broad coverage of standard E6B functions. Sporty’s E6-B provides a defined aviation function set; the CX-3 adds weight-and-balance, holding, and expanded planning utilities within a fixed menu structure. &
Extensive and expandable suite including navigation, performance, atmospheric modelling, and visual analysis tools such as Holding Pattern Computer, Wind/Diversion Visualiser, Carburettor Icing Risk Assessment, and aircraft-specific performance planning. \\
\hline

\textbf{Unit conversions} &
Manual conversion using printed scales; limited to available graduations. &
Automated unit conversion across predefined categories and units. &
Dedicated aviation unit conversion system covering mass, volume, distance, speed, temperature, pressure, and related quantities. \\
\hline

\textbf{Data persistence} &
None by design; all calculations are ephemeral and must be recreated manually. &
Device-dependent. Sporty’s E6-B clears memory on battery removal; the CX-3 provides non-volatile storage for aircraft profiles, trip plans, and weight-and-balance data, with backup and restore support. &
Platform-dependent persistence; supports saved aircraft profiles, weight-and-balance configurations, and progressive learning workflows. \\
\hline

\textbf{Update and extensibility} &
None; static physical artefact. &
Limited. Sporty’s E6-B is effectively fixed-function; the CX-3 supports firmware updates but remains constrained to manufacturer-defined features and interaction models. &
Software-native update cycle; calculators, visualisers, models, and aircraft modules can be added or refined over time across supported platforms. \\
\hline

\textbf{Aircraft-specific modelling} &
Not supported; calculations are generic. &
Partial. The CX-3 supports aircraft profiles for planning and weight-and-balance; Sporty’s E6-B provides weight-and-balance functions without persistent aircraft profiles. &
Explicit aircraft-specific modelling including graphical weight-and-balance tools with CG envelope visualisation, tailored to individual airframes. \\
\hline

\textbf{Visualisation} &
None beyond manual reading of scales and grids. &
Predominantly numeric display; the CX-3 provides a colour LCD with structured separation of inputs and outputs, but no graphical modelling. &
Rich, task-specific visualisation including Holding Pattern geometry, Wind/Diversion vectors, wind components, Carburettor Icing Risk indication, and Weight-and-Balance CG envelopes. \\
\hline

\textbf{Pedagogical scaffolding} &
High conceptual transparency through manual construction, but minimal guidance for error diagnosis or conceptual reinforcement. &
Procedural prompting improves workflow consistency, but explanatory depth and conceptual feedback remain limited. &
Integrated visual explanation combined with embedded monographs, enabling users to see and understand how inputs affect aerodynamic, navigational, and performance outcomes. \\
\hline

\textbf{Typical error profile} &
Interpolation and reading errors, incorrect scale selection, vector misplacement, and compounding error across chained calculations. &
Input, unit, or mode-selection errors; limited display context can contribute to mis-entry or misinterpretation. &
Input errors remain possible, but visual feedback, unit enforcement, and validation reduce ambiguity; risk shifts toward automation complacency if conceptual understanding is not maintained. \\
\hline

\textbf{Power and failure modes} &
No power required; failure limited to physical loss or damage. &
Battery-dependent hardware; functionality lost on power depletion. &
Dependent on host device battery and operating system stability; mitigated by offline capability and multi-platform availability. \\
\hline

\textbf{Examination suitability (manufacturer-stated)} &
Varies by authority and examination regime. &
Explicitly authorised for use during FAA and Canadian aviation knowledge examinations, as stated in manufacturer documentation. &
Not positioned as an examination test aid; intended for educational and planning use outside formal examination contexts. \\
\hline

\textbf{Platform interoperability} &
Not applicable. &
Dedicated single-purpose hardware; no integration with broader digital workflows. &
Runs across iOS, Android, and Windows platforms; supports integration with modern, multi-device pilot workflows. \\
\hline

\end{tabular}
}}
\caption[Structured comparison of flight computer modalities]{\label{tab:flight_computer_detailed_comparison}
Structured comparison of mechanical, electronic, and software-based flight computers across the evaluation dimensions defined in Section~3.1, including computational architecture, functional scope, pedagogical support, extensibility, error characteristics, and manufacturer-stated examination suitability.}
\end{table}

\subsection{Summary of Comparative Findings}

\subsubsection{CRP-1 Whiz-Wheel (Mechanical Flight Computer)}

The CRP-1, and similar mechanical flight computers, remain valued for their reliability and pedagogical clarity, especially in reinforcing core aerodynamic and navigational concepts through manual interaction. Their offline, tactile nature ensures independence from power sources or digital failure modes, making them robust in austere environments. However, their scope is limited, calculations are slow and prone to human error, and they offer no capacity for data storage, reuse, or visualisation. Additionally, the steep learning curve and need for physical dexterity can reduce accessibility for some users, particularly under time pressure or in turbulence.
Unlike digital tools that offer near-instant precision, whiz-wheels like the CRP-1 demand interpretation, and certain values may fall outside the instrument’s physical range \cite{vanRiet2007}.

\subsubsection{Electronic Flight Calculators}
Electronic flight calculators such as Sporty’s Electronic E6-B and the ASA CX-3 inherit the core computational structure of the traditional E-6B  ---  particularly in wind, time--speed--distance, and fuel planning calculations  ---  while re-implementing these functions in fixed, menu-driven form with digital input and display. Devices such as Sporty’s Electronic E6-B and the ASA CX-3 improve numerical precision, calculation speed, and readability, and have consequently seen widespread adoption in training environments where standardisation and examination compliance are prioritised.
As discussed in Section 1.2, these devices are explicitly authorised for use in aviation knowledge examinations and are optimised for predictability and consistency rather than extensibility. Architecturally, they are implemented as dedicated, menu-driven hardware systems with constrained displays and tightly bounded software models. While the ASA CX-3 supports firmware updates and non-volatile storage of selected planning data, its functionality remains limited to a manufacturer-defined feature set and interaction paradigm.
As a result, electronic flight calculators offer limited scope for the introduction of new calculator classes, advanced visualisation techniques, or pedagogically rich explanatory layers. Even where firmware updates are possible in principle, the closed hardware form factor and non-extensible interaction model \cite{ASACX3Manual2017} impose practical constraints on complexity, visual richness, and educational integration. These limitations become increasingly apparent when electronic flight calculators are compared with modern, multi-platform software-based flight computer applications such as E6BJA, which are not bound by the same architectural constraints.

\textbf{Sporty’s Electronic E6-B}
Sporty’s E6-B modernises many of the core whiz-wheel functions in a compact digital form. It performs well for essential flight planning tasks such as wind correction, altitude conversions, and fuel burn estimation. However, the device’s dated UI, small display, and fixed-function architecture limit its long-term flexibility. It is not user-updateable, and lacks support for aircraft-specific profiles or any form of extensibility. While accurate and exam-compliant, its design is more evolutionary than transformative.

\textbf{ASA CX-3}
The ASA CX-3 builds upon the E6-B concept with an improved screen, better key layout, and a more comprehensive function set  ---  including relative humidity, \%MAC calculations, and additional weight and balance tools. It remains a popular tool authorised for use in FAA and Canadian aviation knowledge examinations. Nevertheless, like its predecessor, the CX-3 is a closed system, reliant on disposable batteries, and constrained to its pre-programmed capabilities. Customisation is not supported, and the device lacks interoperability with other digital tools or platforms. Usability studies have shown that ab-initio pilots benefit significantly from systems that provide visual clarity and accessible interfaces, which traditional hardware systems like the CX-3 often fail to deliver \cite{Schwartzentruber2017}.

\subsubsection{Software-based E6BJA Flight Computer}
The E6BJA app represents a more radical departure from the traditional model. As a fully software-based, multi-platform flight computer, it supports a significantly broader range of aviation calculators  ---  covering standard E6B functions alongside advanced modelling, such as carburettor icing risk, the 1976 ISA atmospheric model, and aircraft-specific weight and balance calculations for over 25 aircraft types. Unlike its hardware-based predecessors, E6BJA is offline by design, ensuring full functionality in flight or in low-connectivity regions. Its integration of explanatory monographs and visualisation tools provides strong cognitive and educational support, lowering the barrier to understanding complex operations. Although some initial learning effort is required, the interface is more discoverable than mechanical systems and benefits from modern UX design principles. Its extensibility, portability, and software-based architecture make it uniquely positioned for future integrations, including real-time weather or airspace data.

\section{Discussion}

The comparative evaluation presented in this paper demonstrates that E6BJA represents a significant evolution in pilot-facing computational tools, exemplified by both the E6BJA flight computer and the software-based Landing Pattern Computer. These systems combine the pedagogical transparency of traditional mechanical flight computers with the flexibility, precision, and extensibility of modern software. Their greatest value lies in real-world, non-examination contexts  ---  such as flight planning, dispatch training, and instructor-led briefing  ---  where visualisation, error transparency, and integrated educational support are essential.\newline

Traditional tools, both mechanical and electronic, retain value in specific use cases. For example, whiz-wheels like the CRP-1 are useful for developing mental arithmetic, spatial reasoning, and resilience under constrained or degraded conditions, skills that remain relevant in emergency scenarios or during training for dead reckoning navigation \cite{Volz2016}. Similarly, electronic calculators like the ASA CX-3 offer predictable performance in exam settings and are permitted in high-stakes assessments due to their fixed-function, non-networked hardware design.\newline

Despite offering significantly expanded capabilities, E6BJA is not currently approved for use in formal regulatory examinations, such as FAA knowledge tests. This is due not to any deficiency in accuracy or transparency, but rather to existing mobile device policies, which prohibit general-purpose or internet-capable devices from examination environments, regardless of whether connectivity is disabled. These restrictions apply even to offline-native applications like E6BJA. This policy constraint has been explicitly considered in E6BJA’s design, which targets educational, instructional, and operational planning contexts  ---  not certified exam usage.\newline

Future extensions to E6BJA may include integration with real-time weather and NOTAM services, improved connectivity with EFB platforms, or interoperability with electronic logbook systems. However, such integrations will be carefully balanced with the app’s offline-first design philosophy to preserve its resilience and platform independence.\newline

Looking ahead, E6BJA holds potential not only for individual pilot training but also as a research tool in aviation education and human factors. Its modular architecture allows for the development of experimental calculator modules or instructional aids, making it suitable for classroom deployment, syllabus integration, and even educational product partnerships.\newline

As digital tools become more pervasive in pilot and aviation professional workflows, the challenge will not only be to match the rigour of traditional methods, but to ensure that these tools foster  ---  not erode  ---  the mental discipline, error awareness, and operational resilience foundational to safe airmanship.\newline

\begin{table}[htbp]
\centering
\makebox[\textwidth][c]{%
\scalebox{0.7}{%
\begin{tabular}{p{5.2cm}p{7.0cm}p{7.0cm}}
\hline
\textbf{System} & \textbf{Strengths} & \textbf{Limitations} \\
\hline

\multicolumn{3}{l}{\textbf{Mechanical Flight Computers}} \\
\hline

\textbf{CRP-1 Whiz-Wheel} &
\begin{itemize}
  \item Reliable and fully offline
  \item Excellent for conceptual learning and teaching fundamental principles
\end{itemize}
&
\begin{itemize}
  \item Limited computational scope
  \item Steep learning curve
  \item Prone to human error
  \item Cannot store or reuse data
  \item No visualisation or contextual guidance
\end{itemize}
\\\hline

\multicolumn{3}{l}{\textbf{Electronic Flight Calculators}} \\
\hline

\textbf{Sporty’s Electronic E6-B} &
\begin{itemize}
  \item Covers essential E6B functions
  \item Accurate and exam-compliant
\end{itemize}
&
\begin{itemize}
  \item Outdated user interface
  \item Not updateable or extensible
  \item No aircraft-specific modelling or profiles
  \item Closed, battery-dependent hardware
  \item No integration with digital ecosystems
\end{itemize}
\\
\hline

\textbf{ASA CX-3} &
\begin{itemize}
  \item Broad range of flight planning functions
  \item Improved screen and user interface
  \item Authorised for FAA and Canadian aviation knowledge examinations
\end{itemize}
&
\begin{itemize}
  \item Limited extensibility
  \item Fixed-function hardware
\end{itemize}
\\
\hline

\multicolumn{3}{l}{\textbf{Software-Based Flight Computers}} \\
\hline

\textbf{E6BJA (Pro Edition)} &
\begin{itemize}
  \item Most feature-rich and computationally extensive
  \item Fully offline and multi-platform
  \item Includes ISA model, carburettor icing, and weight-and-balance for 25+ aircraft
  \item Educational monographs and interactive visualisations
  \item Designed using modern UX principles
\end{itemize}
&
\begin{itemize}
  \item Requires some initial learning
  \item Not certified for formal examinations
  \item Dependent on user device quality (screen size, input methods)
\end{itemize}
\\
\hline
\end{tabular}
}}
\caption[Comparison of flight computer modalities]{\label{tab:flight_computer_comparison}
Comparison of mechanical, electronic, and software-based flight computers, highlighting functional capabilities, pedagogical affordances, and practical limitations across device classes.}
\end{table}

\section{Conclusion}

The E6BJA Flight Computer app represents a natural evolution of pilot-facing computational tools  ---  one that honours the legacy of traditional flight computers while advancing their educational and functional potential through modern software design. Unlike mechanical whiz-wheels or closed-system electronic calculators, E6BJA offers an integrated environment that combines computational rigour with user-centred design, visual clarity, and embedded instructional support.\newline

This paper has demonstrated that E6BJA not only replicates the essential capabilities of its mechanical and electronic predecessors, but extends them in meaningful ways: by supporting aircraft-specific modelling, visualising abstract aerodynamic principles, and scaffolding user understanding through embedded monographs. These enhancements make it not simply a tool for calculation, but a platform for learning, revision, and instruction.\newline

Importantly, E6BJA does not seek to replace traditional tools outright. Instead, it positions itself as a complementary resource, which is especially valuable outside the constraints of standardised examination settings. While devices such as the CRP-1 or ASA CX-3 retain relevance in exams or tactile navigation training, E6BJA excels in real-world planning, self-guided study, and instructor-led briefings where flexibility, comprehension, and situational adaptation are essential.\newline

By fusing tradition with innovation, E6BJA illustrates how modern software can enhance  ---  not undermine  ---  the cognitive discipline of flight training. Its development reflects a commitment to safer, smarter, and more informed aviation practice, and opens avenues for further research, collaboration, and pedagogical exploration in the design of pilot-facing technology.

\section*{Disclaimer}

E6BJA is intended for educational and instructional use and is not certified for formal examinations or operational deployment. It complements, rather than replaces, exam approved tools by supporting modern, flexible, and pedagogically grounded flight training for pilots and flight planning professionals.

\end{document}